\documentclass{aa}  

\usepackage{subfigure}
\usepackage{courier}
\usepackage{graphicx}
\usepackage{grffile} 
\usepackage{txfonts}
\usepackage{natbib}
\usepackage[flushleft]{threeparttable}

\begin{document}

   \title{Duty cycle of the radio galaxy B2 0258+35}

   \author{M. Brienza\inst{1,2}\fnmsep\thanks{brienza@ira.inaf.it}
          \and
          R. Morganti\inst{1,2} 
          \and
          M. Murgia\inst{3}
    	  \and
    	  N. Vilchez\inst{4}
    	  \and
          B. Adebahr \inst{1}
		  \and
          E. Carretti\inst{3,7}
          \and
          R. Concu\inst{3}
          \and
          F. Govoni\inst{3}
          \and
          J. Harwood\inst{1} 
          \and
          H. Intema\inst{5}
          \and
          F. Loi\inst{3,6}
          \and
          A. Melis\inst{3}
          \and
          R. Paladino\inst{7}
          \and
          S. Poppi\inst{3}
          \and
          A. Shulevski\inst{1}
          \and
          V. Vacca\inst{3}
          \and
          G. Valente\inst{8}
          }
          
   \institute{ASTRON, the Netherlands Institute for Radio Astronomy, Postbus 2, 7990 AA, Dwingeloo, The Netherlands
            \and
            Kapteyn Astronomical Institute, Rijksuniversiteit Groningen, Landleven 12, 9747 AD Groningen, The Netherlands
		    \and
		    INAF-Osservatorio Astronomico di Cagliari, Via della Scienza 5, I-09047 Selargius (CA), Italy
		    \and
		    Qatar Environment and Energy Research Institute (QEERI), HBKU, Qatar Foundation, PO Box 5825, Doha, Qatar
		    \and
		    Leiden Observatory, Leiden University, Niels Bohrweg 2, 2333 CA Leiden, The Netherlands
			\and
			Dipartimento di Fisica, University of Cagliari, Strada Prov.le Monserrato-Sestu Km 0.700, I-09042 Monserrato (CA), Italy
			\and
			INAF - Istituto di Radioastronomia, Via Piero Gobetti 101, I–40129 Bologna, Italy
			\and
			Agenzia Spaziale Italiana (ASI), Roma}

\abstract
{Radio loud active galactic nuclei (AGN) are episodic in nature, cycling through periods of activity and quiescence. The study of this duty cycle has recently gained new relevance because of the importance of AGN feedback for galaxy evolution.}
{In this work we investigate the duty cycle of the radio galaxy B2~0258+35, which was previously suggested to be a restarted radio galaxy based on its morphology. The radio source consists of a pair of kpc-scale jets embedded in two large-scale lobes ($\sim$240 kpc) with relaxed shape and very low surface brightness, which resemble remnants of a past AGN activity. }
{We have combined new LOFAR data at 145 MHz and new Sardinia Radio Telescope data at 6600 MHz with available WSRT data at 1400 MHz to investigate the spectral properties of the outer lobes and derive their age.}
{Interestingly, the spectrum of both the outer northern and southern lobes is not ultra-steep as expected for an old ageing plasma with spectral index values equal to $\rm \alpha_{1400}^{145}=0.48\pm0.11$ and $\rm \alpha_{6600}^{1400}=0.69\pm0.20$ in the outer northern lobe, and $\rm \alpha_{1400}^{145}=0.73\pm0.07$ in the outer southern lobe. Moreover, despite the wide frequency coverage available for the outer northern lobe (145-6600~MHz), we do not identify a significant spectral curvature (SPC$\simeq$0.2$\pm0.2$).  }
{While mechanisms such as in-situ particle reacceleration, mixing or compression can temporarily play a role in preventing the spectrum from steepening, in no case seem the outer lobes to be compatible with being very old remnants of past activity as previously suggested (with age $\gtrsim$ 80 Myr). We conclude that either the large-scale lobes are still fuelled by the nuclear engine or the jets have switched off no more than a few tens of Myr ago, allowing us to observe both the inner and outer structure simultaneously. Our study shows the importance of combining morphological and spectral properties to reliably classify the evolutionary stage of low surface brightness, diffuse emission that low frequency observations are revealing around a growing number of radio sources.}


   \maketitle


\section{Introduction}

Some of the outstanding questions in the study of radio-loud active galactic nuclei (AGN) are concerned with how long and how frequently they are active, and how to identify the various phases of their evolution. Statistical studies and models suggest that the AGN duty cycle strongly depends on the power of the radio source (\citealp{best2005, shabala2008, turner2015}) but precise estimates are still not available. Understanding the life cycle of AGN jets has recently gained a new broader relevance in the context of radio galaxy evolution \citep{morganti2017}, as they can transfer a significant amount of energy into the ambient medium \citep{mcnamara2007, wagner2011, wagner2012, gaspari2012}. Feedback from AGN is indeed required by all cosmological simulations to explain the quenching of the star formation in early type galaxies and the correlation between galaxy, and black hole properties \citep{dimatteo2005, fabian2012, schaye2015, sijacki2015}. 

Radio observations are particularly suited to trace the different stages of evolution in radio galaxies via the study of their radio morphology and spectra. Evidence of radio galaxies in different phases of their evolution has been presented in the literature as described below.

For example, remnant radio galaxies represent the last phase when the jets have switched off (e.g. \citealp{parma2007, murgia2011, saripalli2012}). 
Recent studies based on low frequency observations and modelling of radio galaxies find fractions of remnants $\lesssim$10-15\%, suggesting a rapid expansion of the plasma in the ambient medium after the nuclear activity has ceased \citep{godfrey2017, brienza2017, mahatma2018}. Observations show that remnants can have different characteristics, probably depending on their stage of evolution. At GHz frequencies they are commonly observed to have ultra-steep spectra ($\rm S\propto \nu^{-\alpha}, \alpha>1.2$, e.g. \citealp{parma2007, hurleywalker2015, shulevski2017}), typical of old ageing plasma, but can also have only moderately steep spectra at MHz frequencies ($\rm 0.6<\alpha<1$, e.g. \citealp{murgia2011, brienza2016, mahatma2018}). Moreover, they can show very weak radio cores indicating that the nuclear activity may not completely switch off but just go through a period of significant suppression \citep{murgia2011, brienza2016, brienza2017, mahatma2018}. 

Sources that exhibit remnant emitting plasma from past activity and, at the same time, new-born jets are named restarted radio galaxies (see \citealp{saikia2009} for a review). These can be effectively used to investigate the duty cycle of the radio AGN. 

The most explicit and well-known signatures of recurrent jet activity on large scales are `double-double radio galaxies' (DDRGs). This class of objects was first defined by \cite{schoenmakers2000} as `radio galaxies consisting of a pair of double radio sources with a common centre' and further investigated by other authors (e.g. \citealp{kaiser2000, konar2013a, orru2015}). For these objects the duration of the quiescent phase between two episodes of jet activity is estimated to be in the range $\rm 10^5-10^7$ yr, and is found to be typically shorter than the duration of the previous active phase equal to $\rm 10^8$ yr on average \citep{konar2013b}. However, the phenomenon of restarted AGN is not limited to DDRGs. Another class of sources is observed to have compact inner jets embedded in large-scale, low-surface brightness lobes (e.g. \citealp{jamrozy2009, kuzmicz2017}). In some of these sources, there are indications that the quiescent phase may be as long as $\rm \geq10^8$ yr \citep{jamrozy2007}. Among the most famous cases of restarted jets is the radio galaxy Centaurus A, where more than one phase of activity has been claimed based on the presence of two distinct morphology structures (\citealp{morganti1999, mckinley2013, mckinley2017}). Moreover, observations of remnant emission on small scales ($<$100 pc) associated to active compact radio galaxies \citep{luo2007, orienti2008} suggest that at the beginning of the jet activity, multiple cycles of short bursts ($\rm 10^3-10^4$ yr) may occur before the jets start to expand to large scales.

The physical mechanisms that drive the radio jets intermittence are still poorly understood and are possibly related to the galaxy environment. Some possible explanations include gas-rich galaxy-galaxy interactions (e.g. \citealp{schoenmakers2000}), perturbations of the accretion process (e.g. \citealp{pringle1997, czerny2009, wu2009}) and chaotic cold accretion (e.g. \citealp{gaspari2013, gaspari2017}). Alternative interpretations to the restarting scenario have also been proposed to explain the observed morphologies of these sources. For example, \cite{baum1990} suggest that compact radio sources can arise within a large-scale radio galaxy if the jet propagation is obstructed or impeded on scales of tens of parsecs to few kiloparsecs. In that case, the extended structure would remain visible along with the confined parsec scale source.

The variety of sources described above suggests that a multitude of situations and evolution histories can occur. As a consequence, the interpretation and classification of restarted sources is not trivial and more examples need to be studied in detail for improving our understanding. In particular, extra constraints can be obtained by studying the spectral properties of these sources on a broad spectral range, from MHz to GHz frequencies, in combination with the source morphology. A better characterization of known restarted radio galaxies is also essential for a better selection and census of restarted sources in upcoming large-area radio surveys (e.g. the LOw Frequency ARray (LOFAR) Two-metre Sky Survey - LoTSS, \citealp{shimwell2017}).

In this paper we present new observations of the radio source B2~0258+35, where two low-surface brightness extended lobes (240 kpc) have been discovered at 1400 MHz \citep{shulevski2012} around a compact radio galaxy (3 kpc). Due to its morphology, this source has been interpreted as a restarted radio galaxy in which the outer lobes represent old remnants of a past AGN activity. The aim of this work is to put new constraints on the physical properties and duty cycle of these lobes by studying their spectral properties and morphology over a wide frequency range.    

The cosmology adopted in this work assumes a flat universe and the following parameters: $\rm H_{0}= 70\  km \ s^{-1}Mpc^{-1}$, $\Omega_{\Lambda}=0.7, \Omega_{M}=0.3$. At the redshift of B2~0258+35 equal to z=0.0165 \citep{wegner1993}, 1~arcsec corresponds to 0.34 kpc. Throughout the paper the spectral index, $\alpha$, is defined using the convention $S\propto \nu^{\alpha}$.

\section{Overview on the source B2 0258+35}
\label{overview}

The radio source B2~0258+35 is hosted by the giant early-type galaxy NGC 1167 ($\rm M_B=-21.7 \ mag, \ D_{25} = 56 \ kpc$), which is optically classified as a Seyfert 2 galaxy \citep{ho1997}. The galaxy is located at redshift z = 0.0165 \citep{wegner1993} and does not belong to any galaxy group but is surrounded by a few smaller satellites \citep{struve2010}. Deep optical observations show a faint, tightly wound spiral structure at $r<$30~kpc, which represents the signature of a past major merger event \citep{emonts2010}. 

Indications of past and/or ongoing interactions are also confirmed by the presence and kinematics of a large ($\sim$160 kpc) and massive HI disk ($\rm M_{HI}=1.5\times10^{10} \ M_\odot$, \citealp{noordermeer2005}, \citealp{emonts2010}, \citealp{struve2010}). This HI structure is thought to have assembled via accretion of gas-rich satellite galaxies. However, the gas distribution is extremely regular out to 65 kpc, with disturbed kinematics only in the very outer part, which is possibly the result of recent interactions with its satellites. Thanks to its inner regular kinematics \cite{struve2010} have computed a tight lower limit on the epoch of the last major merger event equal to $\sim$1 Gyr. This timescale is also supported by the absence of young stellar population signatures in the optical spectrum \citep{emonts2006}. However, more recent interactions with small companion galaxies cannot be ruled out.

HI is also observed in absorption against the the central radio source (\citealp{emonts2010} and Murthy et al. in prep). The broad absorption profile is likely the result of the combination of a circumnuclear disk disturbed by the interaction with the radio jet (Murthy et al. in prep). Support for an on-going interaction also comes from the molecular gas (Murthy et al. in prep using observations from \citealp{prandoni2007}). Surprisingly, there is no evidence of broadening in the optical emission lines of the ionized gas (\citealp{emonts2006}, Santoro priv. comm.).

The central radio source has a luminosity at 408 MHz of $\rm L_{408}=10^{24.37} \ W \ Hz^{-1}$ and has been classified as a Compact Steep Spectrum (CSS) source by \cite{sanghera1995}. As shown in the inset in Fig. 1, sub-arsecond imaging shows two plum-like lobes of $\sim$3 kpc total projectd size, with jet-like structures but without clear hot-spots (\citealp{giroletti2005}, \citealp{giovannini2001}). From the jet to counter-jet ratio at 22000 MHz, \cite{giroletti2005} have constrained the jets inclination angle to a value in the range $\rm 40 \ deg<\theta<50\deg$, which is in accordance with the optical classification of the parent galaxy. Using radiative evolution models they have derived a source age of 0.9 Myr and, based on dynamical arguments, have suggest that the radio source might not grow into an extended FRI/II radio galaxy. This is further supported by the clear bending of the southern lobe indicating a dense surrounding interstellar medium (ISM), which may be the cause of the source confinement and is in agreement with the jet-ISM interaction inferred from HI absorption (Murthy et al. in prep). The spectral index distribution within the CSS source varies from $\alpha_{8.4}^{22}$=0 in the core, to $\alpha_{8.4}^{22}$=0.6 in the jet-like structures within the lobes, to $\alpha_{8.4}^{22}$=1.0-1.5 in the surrounding diffuse emission \citep{giroletti2005}.

\begin{figure}
\centering
{\includegraphics[width=9cm]{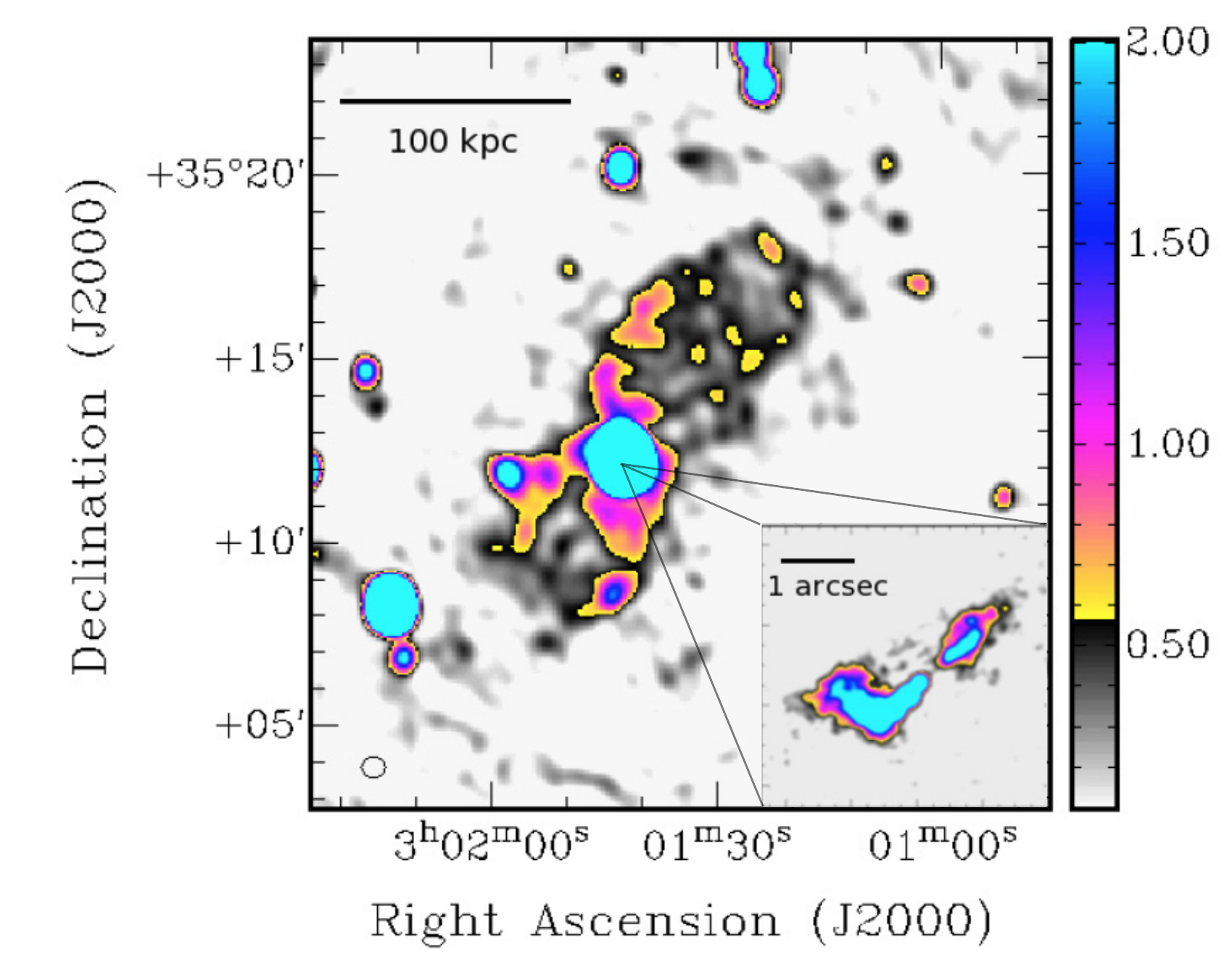}}
\caption{Continuum image of the radio galaxy B2 0258+35 at 1400 MHz obtained with the Westerbork Synhesis Radio Telescope from \cite{shulevski2012}. The synthesized beam is indicated with the ellipse at bottom left. The intensity ranges from 100 $\rm \mu Jy$ (1$\rm \sigma$) to 2 mJy (20$\rm \sigma$). The inset at the bottom right shows a high-resolution radio image at 8400 MHz of the central CSS source obtained with the Very Large Array by \cite{giroletti2005}. }
\label{fig:wsrt_image}
 \end{figure}

A very interesting ingredient for understanding the evolutionary history of the source is the presence of two large-scale (240 kpc), low-surface brightness (1.4~mJy~$\rm arcmin^{-2}$ at 1400 MHz) lobes that surround the CSS source. These have been first detected and studied at 1400 MHz by \cite{shulevski2012} using the Westerbork Synhesis Radio Telescope (WSRT) (Figure \ref{fig:wsrt_image}). The authors propose that the lobes represent old remnants of a previous cycle of jet activity. The S-shape morphology of the broad enhancements in surface brightness located at the edges of the outer lobes is also notable.

Assuming the lobes to be buoyant bubbles expanding in the intergalactic medium \cite{shulevski2012} estimate the dynamical age of the source to be $\gtrsim$80 Myr. Because the age of the young central CSS source is about 0.9 Myr, they suggest that the quiescent phase between the two radio bursts has lasted $\sim$100 Myr.

The main characteristics of the radio source B2~0258+35 and of its host galaxy are summarised in Table \ref{tab:galaxy}.

 \begin{table}[h]

	\small
 \caption{Summary of the properties of the radio source B2~0258+35 and of its host galaxy.}
	\centering
		\begin{tabular}{c  c}
		\hline
		\hline
		Host galaxy & NGC 1167 \\
		$ \rm M_B$ & -21.7 mag\\
		$\rm D_{25}$ & 56 kpc \\
		Optical classification & Seyfert 2\\
		Redshift & 0.0165 \\
		Size inner radio lobes & 3 kpc\\
		Inner lobes radio power & $\rm L_{408}=10^{24.37} \ W \ Hz^{-1}$\\
		Size outer radio lobes & 240 kpc \\
		$\rm M_{HI}$ & $\rm 1.5\times10^{10} \ M_\odot$\\
		Diameter Hi disk & 160 kpc \\ 
 		\hline
		\hline	
		\end{tabular}
   
	\label{tab:galaxy}
\end{table}

\section{Data}

To investigate the radio spectral properties of the extended lobes and to constrain any spectral curvature, we expand here the study at 1400 MHz presented by \cite{shulevski2012} to other frequencies. We have performed observations at 145, 350 and 6600 MHz using the Low-frequency Array (LOFAR, \citealp{vanhaarlem2013}), the Karl G. Jansky Very Large Array (VLA), and the Sardinia Radio Telescope (SRT, \citealp{bolli2015, prandoni2017}) respectively. Moreover, we have reprocessed archival observations of the Giant Metrewave Radio Telescope (GMRT; \citealp{swarup1991}) taken at 235 and 612 MHz. Unfortunately, due to the data quality and dynamic range limitations caused by the bright central source, we could not recover the outer lobe emission at 235, 350 and 612 MHz. However, these frequencies have been useful for the study of the compact source. For the analysis of the outer lobes we have also used the WSRT image at 1400 MHz presented by \cite{shulevski2012}. In the following sections we describe the observations and the data reduction procedures. A summary of the observational details is presented in Table~\ref{tab:data} and a summary of the final image properties is presented in Table \ref{tab:image}.

\begin{table*}[t]

	\small
 \caption{Summary of the observational details. An asterisk * indicates archival observations.}
	\centering
		\begin{tabular}{c  c  c  c  c  c  c}
		\hline
		\hline
		Telescope & Configuration & Frequencies & Target  & Calibrator & Calibrator & Observation date \\
		& & MHz &  TOS (hr) & & TOS (hr) & \\
		\hline
		\hline
		LOFAR & HBA Inner & 118.9-176.9 & 8 & 3C48 & 0.53 & 13 September 2015 \\
		GMRT* & - & 235 & 2.3 & 3C48, 0432+416 &0.75 & 24-25 July 2011\\
		VLA & CnB-B & 224-480 & 2.25 & 3C48 & 0.5 & 6 February - 10 April 2015\\
		GMRT* & - & 612 & 2.3 &  3C48, 0432+416 &0.75 & 24-25 July 2011\\ 
		SRT & - &6000-7200 & 11 & 3C286, 3C138 & 2 & 1-2-6-7 February 2016 \\
		\hline
		\hline	
		\end{tabular}
   
	\label{tab:data}
\end{table*}

\subsection{LOFAR HBA observations and data reduction}

Dedicated observations of the source were performed using the LOFAR High Band Antenna (HBA) on September 13th, 2015. All the 64 antennas of the Dutch Array have been used, providing a maximum baseline of $\sim$100 km. The target was observed in scans of 30 minutes for a total integration time of 8 hours, interleaved by 2-minutes observations of the flux-density calibrator 3C48. The sampling time was set to 1 second and four polarization products (XX, YY, XY, and YX) were recorded. The total observed bandwidth is equal to 95.1 MHz in the range 100.2-195.3 MHz. This was divided in 487 sub-bands of 195.3 kHz composed of 64 channels each. The observational details are summarized in Table~\ref{tab:data}. 

Using the observatory pipeline \citep{heald2010} the data were pre-processed. Data below 108.9 MHz and above 176.9 MHz were flagged and excluded from further processing due to the presence of strong RFIs.

After pre-processing the data were calibrated using the classical direction-indipendent procedure (e.g. \citealp{brienza2016, mahony2016}). Amplitude and phase solutions for each station were calculated according to the model of 3C48 presented by \cite{scaife2012} and transferred to the target field. The data were then combined into groups of ten subbands (corresponding to a bandwidth of $\sim$2 MHz) and phase self-calibrated iteratively, adding progressively longer baselines. The LOFAR imager AWImager \citep{tasse2013} was used for the imaging, which performs both w-projection to account for non-coplanar effects \citep{cornwell1992} and A-projection to account for the changing beam throughout the observation \citep{bhatnagar2008}.

We produced images at different resolutions and with different weightings. The best detection of the outer lobes was achieved with a uv-cut equal to 2 k$\lambda$, robust weighting equal to 0 and final resolution of 80~arcsec~$\times$~98~arcsec. Hints of the lobes emission are observed at higher resolution too but the image quality is not enough for a reliable detection and flux density measurement of the low surface brightness emission due to the presence of deconvolution artefacts associated with the strong central source. The calibration scheme created by \cite{vanweeren2016} and \cite{williams2016}, which corrects for direction-dependent errors due to ionospheric phase and amplitude distortions, did not provide any significant improvement of the dynamic range at resolutions higher than 80~arcsec~$\times$~98~arcsec. 

Therefore, in the following work we have used the image obtained with the original direction-independent calibrated data. The final image was obtained by combining all images at different frequencies in the image plane. It has a central frequency of 145 MHz, a spatial resolution of 80~arcsec~$\times$~98~arcsec and a noise of 3 $\rm mJy \ beam^{-1}$ in the proximity of the target (see Figure \ref{fig:mapsb2}, top panel).

\subsection{SRT observations and data reduction}

The source was observed with the SRT on February 1-2-6-7, 2016 as part of the early science programme `SRT Multi-frequency Observations of Galaxy Clusters' (SMOG - PI M. Murgia, see also \citealp{govoni2017, loi2017, vacca2018}). A region of $\rm 0.7 \deg \ \times \ 0.7 \ deg$ was mapped using the C-Band receiver for a total of 11 hours. 
We performed several on-the-fly (OTF) mappings in the equatorial frame in both right ascension (RA) and declination (DEC). The FWHM of the beam at this frequency is 2.9~arcmin so we set the telescope scanning speed to 6~arcmin~$\rm s^{-1}$ and the scan separation to 0.7~arcmin to properly sample the beam. Full Stokes parameters were recorded with the SARDARA backend (SArdinia Roach2-based Digital Architecture for Radio Astronomy; \citealp{murgia2016, melis2018}). The correlator configuration was set to 1024 frequency channels of approximately 1.46 MHz for a total bandwidth of 1500 MHz. We then set the Local Oscillator to 5900 MHz and we used a filter to select frequency range from 6000 to 7200 MHz, which is relatively free from strong RFIs. The central frequency is thus 6600 MHz and the total bandwidth is 1200 MHz. A summary of the SRT observations is listed in Table \ref{tab:data}.

After flagging RFIs, we performed the bandpass and flux-density calibration using observations of the sources 3C286 and 3C138. The flux density scale was set according to \cite{perley2013}. All data observed during February 7th, were completely discarded due to low quality resulting from bad weather conditions during the observing run. We processed the data using the proprietary Single-dish Spectral-polarimetry Software (SCUBE; \citealp{murgia2016}). We applied the gain-elevation curve correction to account for the gain variation with elevation caused by the telescope structure changes due to the gravitational stress.

We proceeded with fitting the baseline with a second order polynomial and subtracting it from each calibrated scan. To perform the imaging we projected the data in a regular three-dimensional grid with a spatial resolution of 42~arcsec/pixel so that four pixels subtend the beam FWHM. All scans along the two orthogonal axes (RA and DEC) were then stacked together to produce the Stokes I image. In the combination, the individual image cubes were averaged and de-stripped by mixing their Stationary Wavelet Transform (SWT) coefficients as described in \cite{murgia2016}. We then used the higher Signal-to-Noise image cubes obtained from the SWT stacking as a prior model to refine the baseline fit. 

Because the central source is very strong, a high dynamical range was required to be able to detect the extended, low surface brightness lobes in the central region of the map. To achieve this goal we deconvolved the sky image with a beam model pattern following \cite{murgia2016} in order to remove the beam sidelobes. The deconvolution algorithm interactively finds the peak in the image obtained from the SWT stacking of all images and subtracts a fixed gain fraction (typically 0.1) of this point source flux convolved with the re-projected telescope “dirty beam model” from the individual images. In the re-projection, the exact elevation and parallactic angle for each pixel in the unstacked images are used. The residual images were stacked again and the CLEAN continued until convergence. As a final step, CLEAN components at the same position were merged, smoothed with a circular Gaussian beam with FWHM 2.9~arcmin, and then restored back in the residuals image to obtain a CLEANed image. The final image obtained in this way has a noise of 0.9 mJy $\rm beam^{-1}$ and is shown in Figure \ref{fig:mapsb2} (bottom panel).

\subsection{VLA P-band observations and data reduction}

We performed observations of the source with the VLA in two slots on February 6th, 2015 and April 10th, 2015 with the P-band receiver. On both runs, 27 antennas were used, distributed in CnB-B and B configuration in the two different runs respectively. The target and flux density calibrator 3C48 were observed for 1.5 hours and 20 minutes on the first day and 45 minutes and ten minutes on the second day respectively. The sampling time was set to 2 seconds and four polarization products (RR, LL, RL, and LR) were recorded. The total bandwidth, equal to 256 MHz in the range 224-480 MHz, was divided by default in 16 sub-bands of 16 MHz with 128 frequency channels. The observational details are summarized in Table ~\ref{tab:data}.

All datasets were reduced using the following steps. After having applied the flags suggested by the observatory, the data were flagged automatically using the AOFlagger, visually checked and further manually flagged when required. We used the Common Astronomy Software Applications (CASA, version 4.7, \citealp{mcmullin2007}) to perform the calibration in the standard manner and following the guidelines set out in the online tutorial for continuum P-band data\footnote{\url{https://casaguides.nrao.edu/index.php/VLA_Radio_galaxy_3C_129:_P-band_continuum_tutorial-CASA4.7.0}}. The flux scale was set according to \cite{scaife2012}. Phase and amplitude self-calibration was performed. The final image of the field is 2~deg~$\times$~2~deg in size and was made using a Brigg's weighting with a robustness parameter of 0. The image has a resolution of 30~arcsec~$\times$~30~arcsec and central noise of $\rm \sim 1.2 \ mJy \ beam^{-1}$. As already mentioned, due to severe flagging caused by the presence of extensive RFIs and the dynamic range limitations caused by the bright central source, the sensitivity of the data is not sufficient to detect the low surface brightness emission of the outer lobes of the radio galaxy.

\subsection{GMRT observations and data reduction}

We used archival data of the GMRT at 235 and 612 MHz. The observations were performed on July 24th and 25th, 2011. The target was observed in five time-scans for a total integration time of 2.3 hours. The source 3C48 was used as flux-density calibrator and observed at the beginning and at the end of the observing run for ten minutes respectively. We adopt the Scaife-Heald model \citep{scaife2012} for setting the absolute flux scale. Data at both frequencies were recorded simultaneously in single-polarization mode using a 33-MHz bandwidth divided into 512 channels of 65-kHz bandwidth and a sampling time of 16.1 seconds.

We processed the data using the SPAM pipeline \citep{intema2014,intema2017}. The output calibrated visibility data were imported into CASA to produce images at different resolutions. By using uniform weighting during imaging, we obtained a resolution of 3.6~arcsec~$\times$~6.5~arcsec and 9.3~arcsec~$\times$~13.6~arcsec in the 612-MHz map and 235-MHz maps, respectively. The noise is equal to $\rm 200 \ mJy \ beam^{-1}$ at 612 MHz and $\rm 1.2 \ mJy \ beam^{-1}$ at 235 MHz. We also imaged the data using Briggs weighting with robust=0 and uvtaper=40~arcsec to enhance the large-scale emission. The final images with $\sim40$~arcsec resolution have a noise levels equal to $\rm \sim1.2 \ mJy \ beam^{-1}$ and $\rm 12 \ mJy \ beam^{-1}$ at 612 MHz and at 235 MHz, respectively. Unfortunately, with neither imaging weighting schemes we managed to detect the low surface brightness emission of the outer lobes of the radio galaxy.

\begin{table}[t]
\small
\caption{Summary of the image properties at different frequencies. The asterisk indicates the image taken from \cite{shulevski2012}.}
	\centering
		\begin{tabular}{c  c  c}
		\hline
		\hline
		Frequency & FWHM & RMS \\
		MHz &  arcsec$^2$ &  mJy $\rm beam^{-1}$  \\
		\hline
		\hline
		145 & 80$\times$98 & 3 \\
		235 & 9.3$\times$13.6 & 1.2\\
		350 & 30$\times$30 & 1.2\\
		612 & 3.6$\times$6.5 & 0.2\\
		1400* & 33$\times$39 & 0.1\\
		6600 & 174$\times$174 & 0.9\\
		\hline
		\hline	
		\end{tabular}
 \label{tab:image}
\end{table}

\section{Results}
\label{results}

\subsection{Morphology}
\label{morpho}

\begin{figure}[t]
\centering
{\includegraphics[width=9cm]{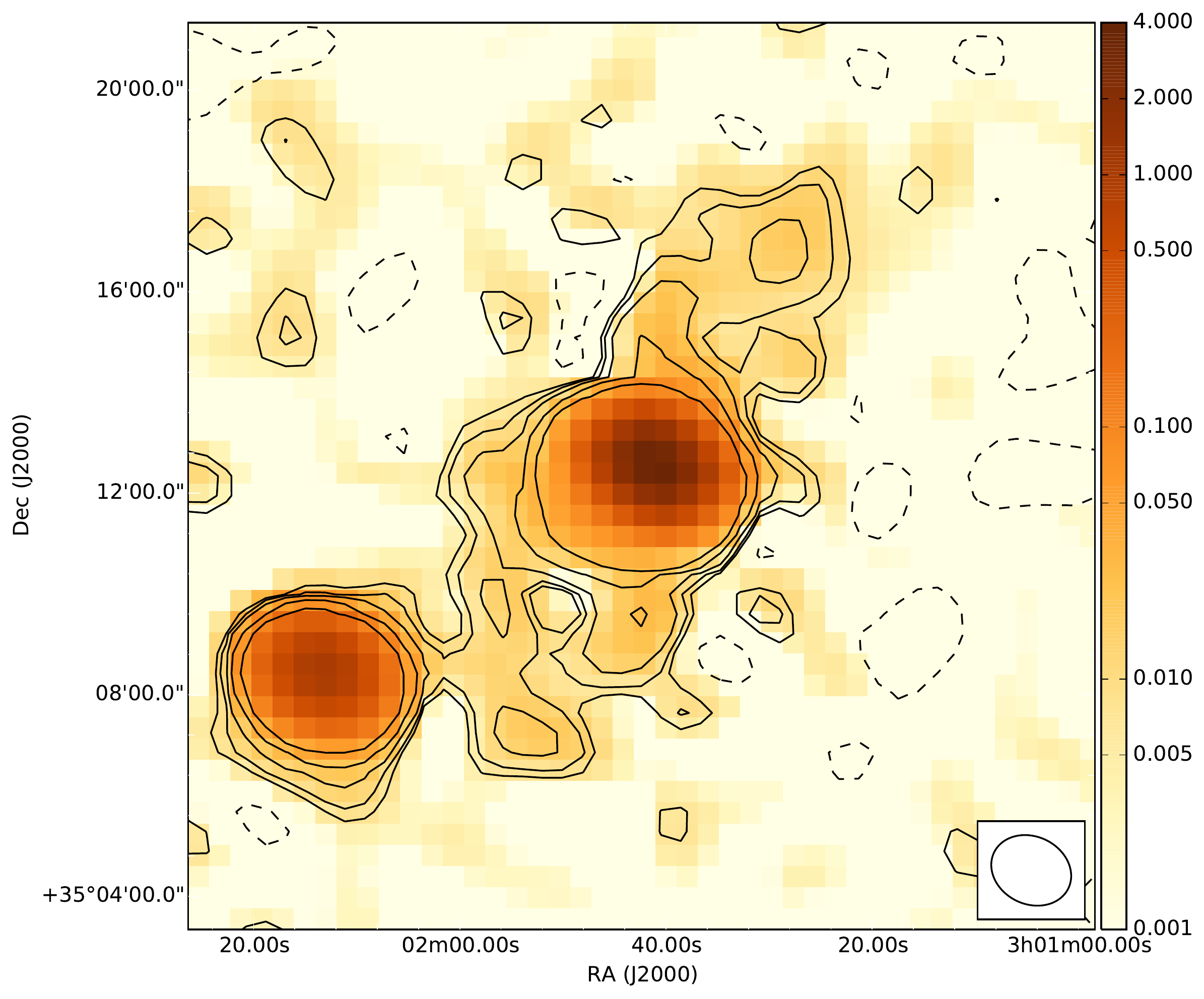}}
{\includegraphics[width=9cm]{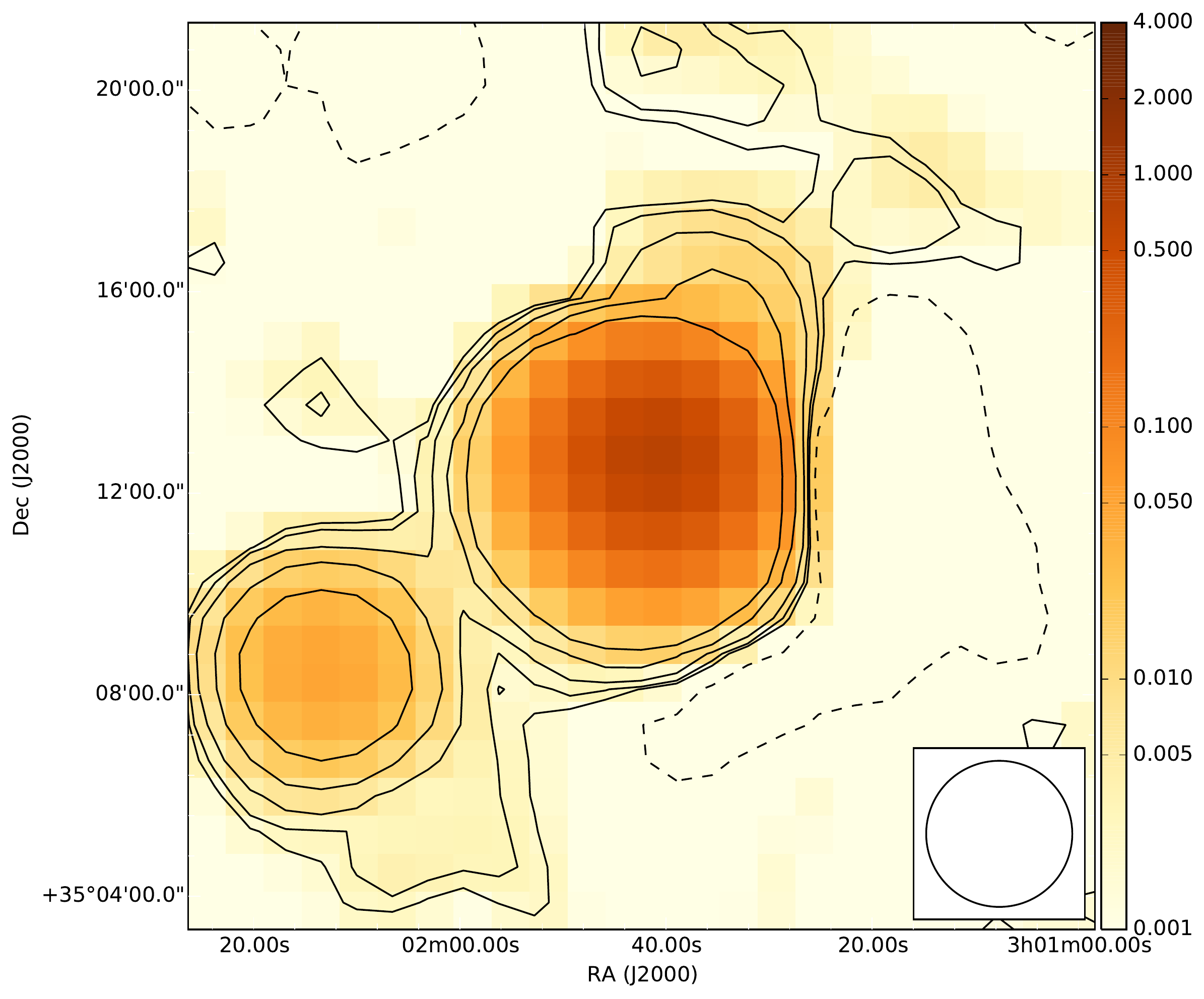}}
\caption{Radio maps of the source B2~0258+35. Top: LOFAR 145-MHz map at 80~arcsec~$\times$~98~arcsec resolution. Levels: -2, 2, 3, 5, 10, 20 $\times \ \sigma$ (3 mJy $\rm beam^{-1})$. Bottom: SRT 6600-MHz map at 2.9-arcmin resolution. Levels: -2, 2, 3, 5, 10, 20 $\times \ \sigma$ (0.9 mJy $ \rm beam^{-1})$. The colour scale is set in $\rm Jy \ beam^{-1}$. In both panels the beam is shown at the bottom right corner.}
\label{fig:mapsb2}
 \end{figure}

In Figure \ref{fig:mapsb2} we show the two radio continuum images of the source B2~0258+35 where the outer lobes are detected at 145 MHz and 6600 MHz respectively. As already mentioned in Section \ref{tab:data}, the low surface brightness emission of the outer lobes has not been recovered at either 235 MHz, 350 MHz or 612 MHz, therefore we do not show those images here. We also note the presence of a point-like unrelated source located in south-east direction (RA 03:02:13.12, DEC +35:08:20.79, J2000).

In the LOFAR map both lobes are clearly visible. Their linear extension and morphology is in general agreement with what has been previously observed at 1400 MHz (240 kpc, \citealp{shulevski2012}). We measure a total angular extension of $\sim$13~arcmin, using the 3$\rm \sigma$ contours at the tip of each lobe as a reference, which corresponds to a linear size equal to $\sim$265 kpc. The observed size difference with respect to \cite{shulevski2012} may be partially attributed to the lower resolution of the LOFAR image and partially to an intrinsic larger extent of the source at low frequencies, especially of the southern lobe.

The total flux density of the outer lobes at 145 MHz is 285 mJy (obtained as the difference between the total flux density of the source and the peak flux density of the central component) and their average surface brightness is equal to 4.7~mJy~arcmin$^{-2}$. We note that the outer lobes represent only $\rm \sim 6$\% of the entire radio emission of the source at 145 MHz. 

Due to the lower angular resolution of the LOFAR map, the S-shape morphology that is clearly visible in the map at 1400 MHz presented by \cite{shulevski2012} (see Figure \ref{fig:wsrt_image}) is here less pronounced. However, we can still recognize two enhancements in surface brightness along the eastern edge of the northern outer lobe and the western edge of the southern outer lobe close to the core.

In the SRT image at 6600 MHz, the northern outer lobe is clearly detected too, although not for its entire extension (see Figure \ref{fig:mapsb2}, bottom panel). Unfortunately, the low resolution of the map does not allow us to study the lobe sub-structures. The southern outer lobe is completely blended with the unrelated source in south-east direction as well with the central CSS source itself, preventing us from any further analysis.

\subsection{Spectral properties}
\label{spectrum}

\subsubsection{Northern outer lobe}

To measure the spectral index of the northern lobe we have convolved the new LOFAR image and the WSRT image produced by \cite{shulevski2012} to the resolution of the SRT map at 6600 MHz equal to 2.9~arcmin. Because the instruments used in this work differ a lot from each other it is worth addressing the problem of the missing flux density. As this is not an issue for the single dish observations, it may become significant when using interferometers. The total flux density of an extended structure in a radio map can indeed be underestimated if the uv-plane does not provide enough uv-coverage at short spacings. An interferometer is sensitive to all the signal which is coming from structures in the sky with sizes $<$0.6$\rm \lambda/D_{min}$, where $\rm D_{min}$ is its shortest baseline \citep{tamhane2015}. The angular size of the northern lobe is $\sim$4.5~arcmin~$\times$~2.5~arcmin. The minimum observed baselines in the different observations are $\rm D_{min,145MHz}\simeq40 \ m$, $\rm D_{min,350MHz}\simeq85 \ m$ and $\rm D_{min,1400MHz}\simeq30 \ m$ giving a largest angular scale ($\theta$, LAS) equal to $\rm \theta_{145MHz}\simeq1.7$ deg, $\rm \theta_{350MHz}\simeq21$~arcmin and $\rm \theta_{1400MHz}\simeq14$~arcmin respectively. We, therefore, should be able to safely detect all the extended emission at all frequencies and making consistent comparisons between single-dish and interferometric data.

We have extracted the flux density from one region drawn in correspondence of the northern lobe (region\textunderscore north) as shown in Figure \ref{fig:regions} (top panel). The measured values are reported in Table \ref{tab:fluxes}. The errors on the flux densities are computed by combining in quadrature the flux scale error and the image noise as shown in \cite{klein2003}. The spectral index analysis in this region reveals a spectrum with very little curvature over the entire frequency range 145-6600 MHz with $\rm \alpha_{1400}^{145}=0.48\pm0.11$ and $\rm \alpha_{6600}^{1400}=0.69\pm0.20$. The spectral curvature (SPC), defined as $\rm \alpha_{6600}^{1400}-\alpha_{1400}^{145}$ \citep{murgia2011}, is equal to SPC$\simeq$0.2$\pm0.2$, well below the typical value of SPC$>$0.5 expected for ageing plasma. Errors on the spectral indices are computed using the following expression:

\begin{equation}
\label{erralpha}
\rm \alpha_{err} = \frac{1}{ln\frac{\nu_1}{\nu_2}}\sqrt{\left(\frac{S_{1,err}}{S_1}\right)^2+\left(\frac{S_{2,err}}{S_2}\right)^2}
\end{equation}

where $S_1$ and $S_2$ are the flux densities at frequencies $\nu_1$ and $\nu_2$ and $S_{1,err}$ and $S_{2,err}$ are the respective errors.

\subsubsection{Southern outer lobe}
\begin{figure}
\centering
{\includegraphics[width=9cm]{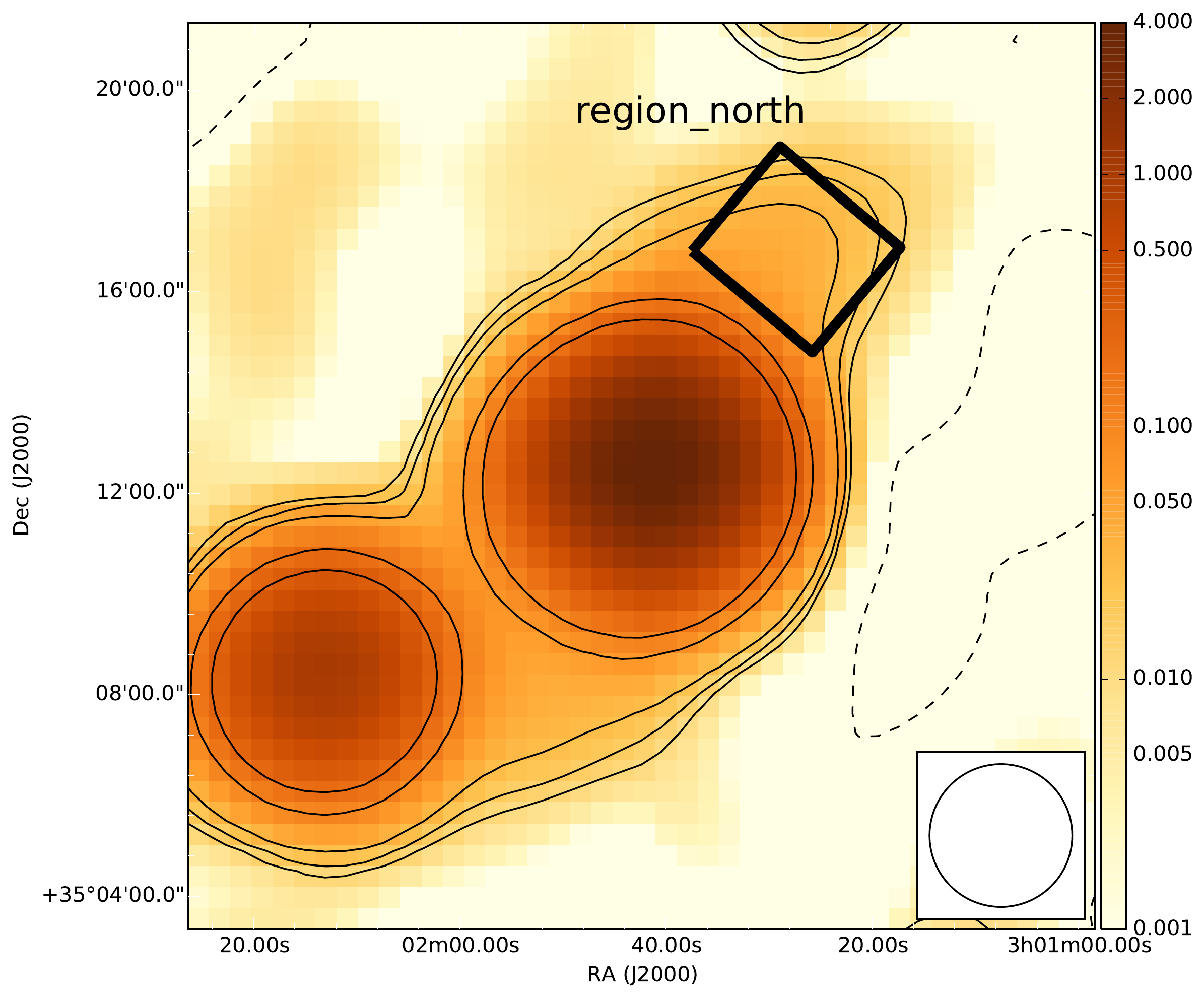}}
{\includegraphics[width=9cm]{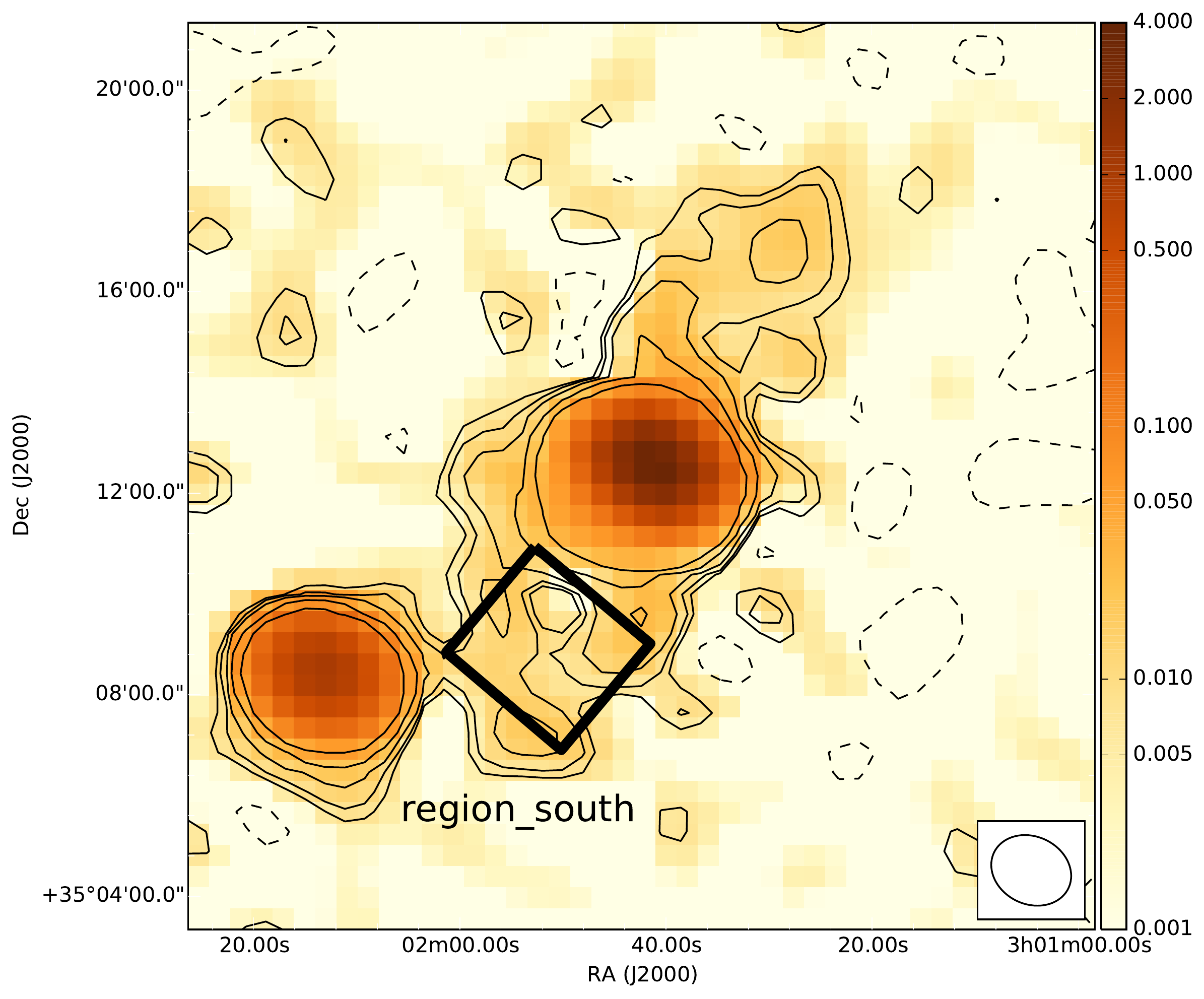}}
\caption{\textit{Top -} LOFAR image at 145 MHz convolved to 2.9-arcmin resolution. Contours represent -2, 2, 3, 5, 10, 20 $\times \ \sigma$ (5 mJy $\rm beam^{-1}$). \textit{Bottom~-} LOFAR image at 145 MHz and 80~arcsec~$\times$~98~arcsec resolution. Contours represent -2, 2, 3, 5, 10, 20 $\times \ \sigma$ (3 mJy $\rm beam^{-1}$). The black regions represent the boxes used to extract the flux density in the northern lobe and southern lobe respectively. The colour scale is set in Jy $\rm beam^{-1}$.}
\label{fig:regions}
\end{figure}

A measure of the flux density of the southern lobe is only possible at 145 MHz and at 1400 MHz. To make this measurement we have convolved the WSRT image to the resolution of the LOFAR image of 80~arcsec~$\times$~98~arcsec. We have extracted the flux density from one region drawn in correspondence of the southern lobe as shown in Figure \ref{fig:regions} (bottom panel, region\textunderscore south). The measured values are reported in Table \ref{tab:fluxes}. The spectral index is $\rm \alpha_{1400}^{145}=0.73\pm0.07$ and is reported in Table~\ref{tab:alpha}.

\begin{table}[h!]
\small
\caption{Flux densities of the northern lobe extracted from the radio images at 2.9-arcmin resolution and of the southern lobe extracted from the radio images at 80~arcsec~$\times$~98~arcsec resolution. Region\textunderscore north and region\textunderscore south are shown in Figure \ref{fig:regions}.}
	\centering
		\begin{tabular}{c c c}
		\hline
		\hline
		Frequency &  Flux density  & Flux density \\
 		 & region\textunderscore north & region\textunderscore south\\
 		(MHz)  & (mJy) & (mJy)\\
 		\hline
		\hline
		 145 & 26.4$\pm$5.6& 68.1$\pm$8.0\\
	     1400 & 9.0$\pm$1.5& 13.4$\pm$1.3\\
	     6600 & 3.1$\pm$0.9& -\\
		\hline
		\hline	
		\end{tabular}
   	\label{tab:fluxes}
\end{table}

\begin{table}[h!]
\small
 \caption{Spectral indices computed for the northern lobe using region\textunderscore north and for the southern lobe using region\textunderscore south (see Table \ref{tab:fluxes} and Figure \ref{fig:regions}).}
	\centering
		\begin{tabular}{c c c}
		\hline
		\hline
		Spectral index &  region\textunderscore north & region\textunderscore south  \\
		\hline
		\hline
		$\alpha_{1400}^{145}$   & 0.48$\pm$0.11 &0.73$\pm$0.07  \\
	    \\ 
	    $\alpha_{6600}^{1400}$  & 0.69$\pm$0.20 &-  \\
   	    \\
	    $\alpha_{6600}^{145}$   & 0.56$\pm$0.09&-  \\
		\hline
		\hline	
		\end{tabular}
   	\label{tab:alpha}
\end{table}

\subsection{Energetics and age of the outer radio lobes}
\label{energetics}

Using the observed properties of the outer lobe radio emission at different frequencies we can, to first order, evaluate some physical parameters of the radio plasma. 

Assuming equipartition conditions between particles and magnetic field, we have computed an average magnetic field value over the entire extension of the two outer lobes equal to $\rm B_{eq}=1 \ \mu G$ \citep{worrall2006}. 

For this calculation we have assumed a power-law particle distribution of the form $N(\gamma)\propto\gamma^{-p}$ between a minimum and maximum Lorentz factor of $\gamma_{min}=10$ and $\gamma_{max}$=$10^6$, with $p$ being the particle energy power index. We have calculated the volume of the source assuming a cylindrical geometry in the plane of the sky. The radius and height of the cylinder were measured using the 3$\sigma$ contours as a reference in correspondence of the lobe maximum extension and are equal to $R=2.5$~arcmin and $h=12$~arcmin respectively. We have assumed that the particle energy content of the source to be equally distributed between heavy particles and electrons so that their ratio $k=1$, and we have set $p=2.0$ according to the observed low-frequency spectral shape where radiative energy losses are less significant. A value of $S_{1400}=120$ mJy is used following \cite{shulevski2012}. Moreover, this computation assumes that the magnetic field is uniformly distributed in the lobe volume.  

By using the magnetic field $\rm B_{eq}$ calculated in this way and a classical radiative ageing model (\citealp{kellerman1964}; \citealp{pacholczyc1970}) it is possible to get a first order estimate of the spectral age $t_s$ of the particle population in the outer lobes. Indeed the age of the emitting particles remains encoded in the curvature of their radio spectrum. In particular, as the particles age with time, the spectrum gets progressively steeper at frequencies higher than a critical break frequency $\nu_b$, due to preferential radiative cooling of high-energy particles.
As already mentioned in Section \ref{spectrum}, the spectrum of the northern lobe of the source B2~0258+35 only shows a little curvature over the entire observed frequency range 145-6600 MHz. This prevents us from properly modelling it to get a radiative age estimate. However, we can set a conservative upper limit to the source radiative age by assuming a lower limit to the spectral break equal to 1400 MHz, which corresponds to the highest frequency after which the spectrum starts showing a curvature. For this we use the following equation \citep{kardashev1962, murgia2011}:

\begin{equation}
t_s=1590\frac{B_{\rm eq}^{\rm 0.5}}{(B_{\rm eq}^2+B_{\rm CMB}^2)\sqrt{\nu_{\rm b}(1+z)}} \\ ,
\label{eqtime}
\end{equation}

\noindent where $t_s$ is in Myr, the magnetic field $B_{eq}$ and inverse Compton equivalent field $B_{CMB}$ are in $\rm \mu G$ and the break frequency $\rm \nu_{\rm b}$ is in GHz. We note that with such a low value of $B_{eq}$, the radiative cooling of the plasma is dominated by inverse Compton scattering of cosmic microwave background (CMB) photons. The CMB equivalent magnetic field equal to $\rm B_{CMB}=3.25\times(1+z)^2$ has a value of $\rm B_{CMB}=3.36 \ \mu G$. In this way we get an upper limit on the lobe age equal to $\lesssim$110 Myr. The upper limit value of the radiative age is expected to decrease if the magnetic field was here underestimated due, for example, to an excess of non-radiative particles with respect to electrons. This occurrence is considered typical of low power jets, where entrainment of massive particles from the external IGM is dominant due to turbulence \citep{croston2008, massaglia2016}.We also note that by using an alternative derivation of the magnetic field by \cite{beck2005} the value lowers by almost a factor three, and this would cause the upper limit of the radiative age to increase. Finally, it should be kept in mind that some mechanisms, such as in situ particle re-acceleration, compression and mixing may affect this result as we discuss in Section \ref{discussion}.

\subsection{The central compact source}
\label{css}

The flux densities of the central compact source measured from our new images at different frequencies are listed in Table \ref{tab:literature_flux} and marked with an asterisk. The measurements have been performed on images having the same resolution at all frequencies (equal to 2.9~arcmin) by fitting a Gaussian function to the unresolved component. Flux densities at different frequencies collected from the literature using the NASA/IPAC Extragalactic Database are also listed in Table \ref{tab:literature_flux}.  

\begin{figure}
\centering
{\includegraphics[width=0.5\textwidth]{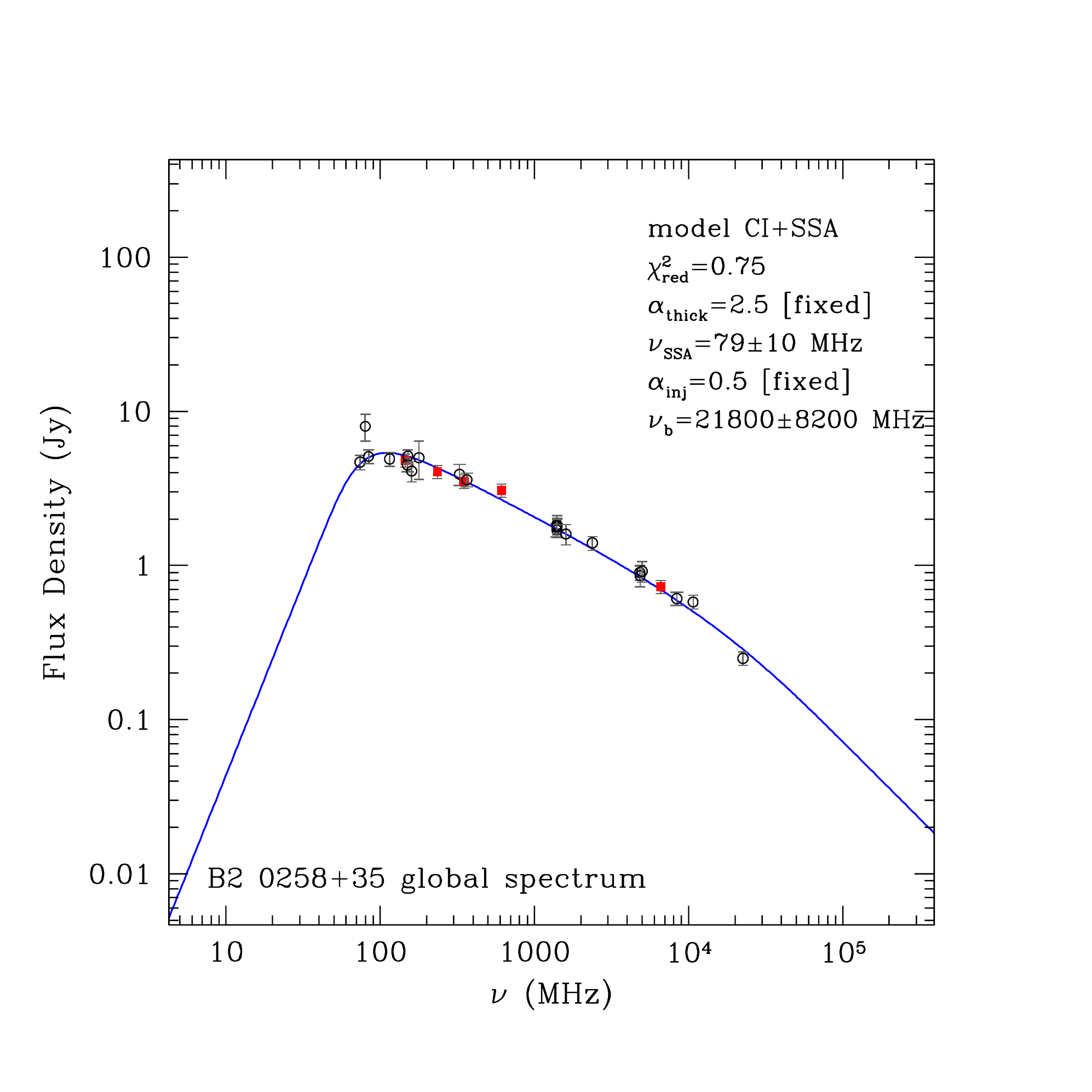}}
\caption{Integrated radio spectrum of the central CSS source and best fit model. Red squares represent the flux density measurements presented in this paper, while black open circles are taken from literature. The entire list of flux densities used in the plot with respective errors and references is presented in Table \ref{tab:literature_flux}.}
\label{fig:css_spec}
 \end{figure}

Thanks to the increased number of flux density measurements with respect to the work of \cite{giroletti2005}, especially at low frequency, we can now better recognize a spectral turnover between 100 and 200 MHz and a possible spectral break at frequencies higher than 5000 MHz (Figure \ref{fig:css_spec}). We use the empirical relation between size and turnover frequency found by \cite{odea1997} to measure the expected turnover frequency. This correlation has been for long explained in terms of synchrotron-self absorption related to the small size of the radio source (e.g. \citealp{snellen2000}, \citealp{fanti2009}). We note, however, that free-free absorption due to a dense ambient medium is also thought to play a role according to some observations (e.g. \citealp{callingham2015, tingay2015}) and simulations \citep{bicknell2018}. Under these circustances the following computed numbers must be treated with care. 

Using the correlation presented in \cite{orienti2014} and a measured linear size for the CSS source equal to 3 kpc we compute a predicted turnover frequency equal to 320 MHz. However, if we deproject the linear size of the source using the inclination angle of 45 deg estimated by \cite{giroletti2005} we get a size of 4.3 kpc and a corresponding turnover frequency equal to 260 MHz. Considered the large scatter present in the correlation (up to almost an order of magnitude), the result is consistent with the observations. 

In order to better quantify the shape of the radio spectrum we have performed a spectral fitting using the software SYNAGE \citep{murgia1999}. In particular, we have used the continuous injection spectral model $S_{CI}(\nu)$ (\citealp{kardashev1962}) modified by low-frequency synchrotron-self absorption \citep{pacholczyc1970} presented in \cite{murgia1999} and shown below:

\begin{equation}
S(\nu)\propto (\nu/\nu_1)^{\alpha + \beta}(1-e^{{-(\nu/\nu_1)}^{-(\alpha + \beta)}})\cdot S_{CI}(\nu),
\label{eq_ssa}
\end{equation}

where $\nu_1$ is the frequency at which the optical depth is equal to 1, $\alpha$ is the spectral index in the transparent frequency range and $\beta$ is the coefficient for an homogeneous synchrotron self-absorbed source and is fixed to 2.5. For the fitting the flux density errors at all frequencies have been set to a systematic value equal to 10\%. We fix the injection index $\alpha_{inj}$ to a classical value of 0.5 and we find a break frequency equal to $\nu_{b}$=21800$\pm$8200 MHz and a synchrotron self-absorption frequency equal to $\nu_{SSA}$=79$\pm$10 MHz. For comparison, we have also fitted the model setting the injection index $\alpha_{inj}$ as a free parameter. In this case we find a break frequency equal to $\nu_{b}$=6400$\pm$3700 MHz and a synchrotron self-absorption frequency equal to $\nu_{SSA}$=65$\pm$24 MHz. We have used the statistical F-test to select the best model and find that the model with free $\alpha_{inj}$ does not improve the fitting with a confidence level of 95\%. Therefore, we consider the model with fixed $\alpha_{inj}$ as the best fit model and we will refer to its results in the following discussion. The integrated spectrum of the central CSS source and the results from the best model fitting are presented in Figure \ref{fig:css_spec}. Our new measurements, indicated as red squares in the figure, nicely follow the literature points, which are shown as black open circles.

The resulting $\nu_{b}$=21800$\pm$8200 MHz is higher than the value estimated by \cite{giroletti2005} equal to 4600 MHz. Assuming a magnetic field equal to $B_{eq}$=90 $\mu G$ \citep{giroletti2005} and using Equation \ref{eqtime}, we get an estimate of the particle radiative age equal to 0.4 Myr. Despite the updated age for the CSS source is lower than the previous estimate by \cite{giroletti2005} (0.9 Myr), it remains high when compared with sources of similar size studied in the literature \citep{murgia2003}. This may support a scenario where the expansion of the lobes is impeded by a dense interstellar medium as suggested by \cite{giroletti2005} and further explored by the study of the HI gas by Murthy et al. in prep.

\begin{table}[h!]
\small
 \caption{Flux densities of the central CSS source at different frequencies with respective references. An asterisk indicates the observations presented in this work.}
	\centering
		\begin{tabular}{l l l}
		\hline
		\hline
		Frequency &  Flux & Reference \\
		(MHz)  	& (Jy) &\\	
		\hline
		\hline
		74    	& 4.69$\pm$0.51 & \cite{cohen2007}\\
		80 		& 8.00$\pm$1.60 & \cite{slee1995}\\
		84 		& 5.10$\pm$0.30 & \cite{hurleywalker2017}\\
		115 		& 4.90$\pm$0.10 & \cite{hurleywalker2017}\\
		145*   	& 4.82$\pm$0.48 & this work\\
		150  	& 4.50$\pm$0.10 & \cite{hurleywalker2017} \\
        151 		& 5.11$\pm$0.17 & \cite{hales1993}  \\
	    160 		& 4.10$\pm$0.60 & \cite{slee1995}\\
		178 		& 5.00$\pm$1.40 & \cite{pilkington1965} \\
		235*   	& 4.06$\pm$0.20 & this work\\
		327 		& 3.91$\pm$0.60 & \cite{rengelink1997}\\
		350*  	& 3.51$\pm$0.18 & this work\\
		365 		& 3.60$\pm$0.04 & \cite{douglas1996} \\
		612*   	& 3.07$\pm$0.15 & this work\\
		1400   	& 1.70$\pm$0.10 & \cite{brown2011}\\
		1400    & 1.77$\pm$0.26 & \cite{white1992} \\
		1400    & 1.84$\pm$0.27 & \cite{condon1998}\\
		1400    & 1.80$\pm$0.18 & \cite{shulevski2012} \\
		1600    & 1.60$\pm$0.24 & \cite{sanghera1995}\\
		2380    & 1.40$\pm$0.07 & \cite{dressel1978} \\
		4835    & 0.90$\pm$0.10 & \cite{griffith1990}\\
		4850    & 0.86$\pm$0.13 & \cite{gregory1991}\\
		5000    & 0.92$\pm$0.14 & \cite{sanghera1995}\\
		6600*   & 0.73$\pm$0.07 & this work\\
		8400    & 0.61$\pm$0.05 & \cite{giroletti2005}\\
		10700   & 0.58$\pm$0.03 & \cite{kellerman1973}\\
		22500   & 0.250$\pm$0.005 & \cite{giroletti2005}	\\
		\hline
		\hline
		\end{tabular}
   	\label{tab:literature_flux}
\end{table}

\section{Discussion}
\label{discussion}

Due to their very low surface brightness and amorphous morphology, the outer lobes of B2~0258+35 have been previously suggested to be old remnants of a past episode of AGN activity. The increase in frequency coverage presented in this work allows us to make a step forward in the understanding of the nature of these structures and in the overall radio galaxy evolution as we describe below.

\subsection{Source characteristics}

Our new LOFAR observations at 145 MHz show that the outer lobes of B2 0258+35 have a total linear extension of 265 kpc and a relaxed shape with low surface brightness (4.7~mJy~arcmin$^{-2}$ at 145~MHz) in agreement with the previous study at 1400 MHz \citep{shulevski2012}. The surface brightness values are consistent with other observations of remnant radio lobes (e.g. 4~mJy~arcmin$^{-2}$, \citealp{brienza2016}; 2.5-6~mJy~arcmin$^{-2}$, \citealp{saripalli2012}), a few giant radio galaxy lobes (e.g. 1 mJy arcmin$^{-2}$, \citealp{saripalli2012}; 10 mJy arcmin$^{-2}$, \citealp{subrahmanyan2006}) and the radio galaxy NGC 3998 (5~mJy~arcmin$^{-2}$, \citealp{frank2016}). The outer lobes are so faint that they only represent $\rm \sim 6$\% of the entire source luminosity. This extreme contrast in surface brightness between the inner and outer structure represents, to date, a quite rare case among the radio galaxy population as further discussed is Section \ref{comparison}. Because the host galaxy is located in a low density environment, the plasma of the outer lobes may have suffered a severe adiabatic expansion and a consequent surface brightness decrease. However, if this is the case, we would expect the outer lobes to dissipate very quickly in the ambient medium. High-sensitivity X-ray observations would be needed to probe the pressure balance between the lobes and the surrounding IGM. At the same time, if the inner jets have strong interactions with the surrounding ISM, the radio luminosity may be boosted due to compression of the magnetic field and the increased particle density \citep{morganti2011, tadhunter2011, giroletti2005}. 

Other features that are interesting to discuss and may provide hints for the interpretation of the source are the broad enhancements in surface brightness present at the edges of the outer lobes. One possible explanation is that this S-shape structure represents an active channel of plasma connecting the inner and outer lobes. Alternatively, it could represent remnant large-scale jets from the previous cycle of activity. The reason for the bent geometry of these features remains unclear though. One possibility is that the jets are bent as a result of the strong interaction with the dense nuclear ISM. Otherwise the jets may have been subject to precession. Precession of radio jets have been observed in radio galaxies (e.g. \citealp{falceta2010}; \citealp{gong2011}, \citealp{nawaz2016}). Models show that on such large spatial scales precession can only happen on timescales larger than 1-10 Myr (e.g. \citealp{nixon2013}), which would be compatible with the timescales we have estimated for the radio galaxy B2 0258+35. The origin of the phenomenon is typically connected to gas misalignments in the accretion disk, which are often a consequence of a merger event, although it is still not completely clear how the timescales of these instabilities and the precession of kiloparsec-scale jets relate (e.g. \citealp{lu2005}). Whether the source B2~0258+35 has actually experienced accretion instabilities is difficult to assess. \cite{shulevski2012} investigate extensively different scenarios for the black hole gas supply and triggering based on the results of the gas kinematics of \cite{struve2010}, \cite{emonts2006} and \cite{prandoni2007}. From that, they exclude a direct link between the merger history of the source and the duty cycle of the radio activity. They suggest instead that the black hole is accreting gas that is cooling from the galaxy's hot corona via Bondi accretion \citep{bondi1952, hardcastle2007}. In this case, chaotic accretion of cold gas clouds condensed from the hot atmosphere may also take place as suggested by a few authors (e.g. \citealp{soker2009, pizzolato2010, gaspari2013, gaspari2017}). This may be responsible for the accretion instabilities required by precession models, as well as they could justify the jet time discontinuity. Moreover, we cannot exclude that some interactions with the small gas-rich satellites that surround the host galaxy may have disturbed the accretion process and influenced the AGN duty cycle.

Beside the morphology, another interesting property of the outer lobes is the spectral shape. We find that the radio spectrum of neither the northern lobe nor the southern lobe is ultra-steep ($\rm \alpha>1.2$), as expected for an old ageing plasma. Moreover, despite the wide frequency coverage available for the northern lobe (145-6600~MHz), no significant spectral curvature is identified  (see Section \ref{spectrum}). Based on the spectral slope and using a simple radiative model, we have derived a first order upper limit to the plasma age in the northern lobe equal to $<$110~Myr (see Section \ref{energetics}). We stress that if mechanisms such as particle re-acceleration, mixing or compression are in action, this age estimate does not hold anymore as it is further discussed in Section \ref{scenarios}). By treating the lobes as buoyant bubbles expanding in the intergalactic medium (IGM) with a rising speed of $\rm \sim$~613$\rm \ km \ s^{-1}$ \cite{shulevski2012} have estimated that the time elapsed since the jet activity has ceased is equal to $\gtrsim$80 Myr (the lower limit is introduced to take into account a possible inclination angle). This value, however, can also be interpreted as an upper limit to the entire age of the outer lobes if we consider that the outer lobes must have expanded to the current size already during the jet activity. In this case the lobes, pushed by the jets advance, should have expanded faster than what assumed by the buoyancy scenario. With this assumption the dynamical age of the outer lobes is consistent with the radiative age upper limit computed in this work.

\subsection{Source evolutionary scenarios and duty cycle}
\label{scenarios}

The characteristics of the source B2~0258+35 are exceptional and a unique interpretation of the source's evolution remains difficult to assess. Here we propose three evolutionary scenarios to explain the spectral and morphological properties discussed above.  

\begin{enumerate}

\item The radio AGN has never switched off. Instead, the large-scale jets have been temporarily disrupted and smothered by e.g. the interaction of the particle flows with a dense medium as suggested by \cite{baum1990}. In this case the two enhancements in surface brightness visible at the edges of the outer lobes could represent two uncollimated jets that connect the outer and inner structure. Under this circumstance the outer lobes would still be fuelled with fresh particles, even if at a low rate, and this would explain the small spectral curvature. This scenario may be consistent with the idea of the inner lobes currently being confined or impeded by a dense interstellar medium as suggested by \cite{giroletti2005} and Murty et al in prep.\\

\item The radio AGN has actually stopped and restarted after a short amount of time. The outer lobes could be currently disconnected from the inner jets and therefore not replenished with fresh particles. However, if the time elapsed since the AGN switch off is not too long, it is possible that the spectral steepening produced by particle ageing in the outer lobes is still not prominent in the considered spectral range (145-6600 MHz). The low level of magnetic field observed would also contribute to make the spectral evolution slow. In this case the two enhancements in surface brightness discussed above could be interpreted as the channels of the previously active jets that have recently stopped to be fuelled. Assuming this scenario, we can use the timescales presented in Section \ref{energetics} to place an upper limit to the time elapsed between the two phases of jet activity equal to $\lesssim 100$ Myr, by combining the radiative age estimate for the outer lobes (<110 Myr, see Section \ref{energetics}) with the CSS radiative age equal to 0.4 Myr (see Section \ref{css}). This is in agreement with the initial estimate by \cite{shulevski2012}. Higher frequency observations would be needed to set tighter constraints on the radiative age of the outer lobes and therefore on the duty cycle. Unfortunately such high sensitivity observations on such large spatial scales are difficult to obtain.\\

\item The outer lobes are remnants of a previous AGN activity and are detached from the current nuclear activity but they do not show an important spectral curvature because of mechanisms such as in situ particle reacceleration, adiabatic compression and particle mixing. All these mechanisms represent valid alternatives to explain the small spectral curvature as they influence the spectral evolution. We stress that, in this case, the computation of the radiative age presented above is not valid. In particular, particle re-acceleration tend to increase the energy of high energy particles causing a `flattening' of the high-frequency tail of the observed radio spectrum \citep{alexandar1987}. This phenomenon may be caused by shocks, like backflows in lobes of active radio galaxies, or stochastic processes, like plasma turbulence. In this occurrences we would also expect significant mixing of particles of different ages within the outer lobes and this would contribute to erase the spectral curvature \citep{turner2017}. By increasing the kinetic energy of the particle population and thus shifting the break frequency to higher frequencies, adiabatic compression may also have a role in contaminating the radiative age estimate \citep{scheuer1968, alexandar1987}. In this case \cite{alexandar1987} suggests that some brightness enhancement should be observed together with some polarized emission produced by the compressed magnetic plasma \citep{laing1980}.

For the source B2~0258+35, re-acceleration and particle mixing may be compatible with the presence of turbulence created as a consequence of the shock evolution, between the low-power jets and the ambient medium \citep{croston2008, massaglia2016}. Adiabatic compression also represents an interesting possibility. As mentioned in Section \ref{morpho}, the outer lobes show at both 145 and 1400 MHz two surface brightness enhancements at the edge of each lobe in specular position with respect to the core that could fit the expected features for adiabatic compression. Polarization information will be used to test this hypothesis (Adebahr et al. in prep). 

\end{enumerate}

Discriminating among the above-mentioned scenarios and providing tight constrain on the duty cycle of B2~0258+35 is difficult. However, our results seem to suggest that in no case the outer lobes are compatible with very old remnants as the morphology alone would suggest and as it was previously proposed. Indeed, all of the mechanisms considered above to explain the small spectral curvature become irrelevant after some time after the large scale jets have switched off. For example \cite{eilek2014} estimates that, for the case of Centaurus A, the decay time of the turbulence is of the order of few tens of Myr after which the lobes should quickly fade away. By assuming comparable values of turbulence decay for B2~0258+35, it follows that, either the lobes are still fuelled by the nuclear activity or the switch off must have happened only few tens of Myr ago, suggesting a quiescent phase shorter than a few tens of Myr, contrary to previous expectations.

\subsection{Comparison with other radio galaxies}
\label{comparison}

As already mentioned above, the properties of B2 0258+35 are peculiar among radio galaxies. However, we have recognized a small number of objects that show some comparable characteristics as described below. 

A few sources, for example, show some inner and outer lobes with a very high surface brightness contrast as observed in B2 0258+35 and for this reason are claimed to be restarted radio galaxies. Among these there are the radio galaxy associated with NGC~3998 ($\sim$20 kpc, \citealp{frank2016}), the source 4C 29.30 ($\sim$600 kpc, \citealp{jamrozy2007}), the misaligned DDRG 3C293 (190 kpc, \citealp{joshi2012, machalski2016}), the source Mrk 6 (7.5 kpc, \citealp{kharb2006}) and the well-known radio galaxy Centaurus A ($\sim$200 kpc, \citealp{mckinley2017, morganti1999}). Moreover, we note some similarity with the famous radio galaxies M87 (e.g. \citealp{owen2000}) and 3C317 \citep{zhao1993, venturi2004}. 
Particularly interesting is the halo of the source M87, which seems to be made of a superposition of many plasma bubbles of different ages and does not show any significant spectral curvature in its spectrum, with no clear relation between the spectral index and the brightness distribution at low frequencies \citep{degasperin2012}. 
It is interesting to note that some of these sources, that is, NGC~3998, Mrk 6 and Centaurus A show a similar S-shape morphology to B2~0258+35. The origin of this shape has been suggested to be related to a change of angular momentum in the accreting gas or to an episodically powered precessing jet \citep{frank2016, kharb2006}.

From a spectral point of view we find a striking resemblance between B2~0258+35 and Centaurus A. Indeed, the radio spectrum of the northern outer lobe of Centaurus A is not ultra-steep up to 90000 MHz (\citealp{hardcastle2009, alvarez2000}) and its age is estimated to range between few tens of Myr (using radiative models, \citealp{hardcastle2009}) and >100 Myr (using dynamical models, \citealp{saxton2001}).
While, in situ particle reacceleration is claimed to be a key component for preventing the spectrum from steepening at high frequency, the discovery of an intermediate-scale northern middle lobe (\citealp{morganti1999,mckinley2013,mckinley2017}) suggests the presence of an `open channel' of fresh particles between the inner and the outer lobe.

While all the afore-mentioned radio galaxies show signs of recurrent jet activity, the origin and timescales of this intermittence are still unclear. The sources 3C317 and M87 are located at the centre of massive clusters of galaxies so they are most likely going through a short, self-regulating AGN feedback cycle with the surrounding hot medium.

All the other sources presented above, including B2~0258+35, are located instead in low density environments. Therefore, their jet activity is expected to be triggered less frequently. It is worth noting though, that they all show signatures of past merger events and a large reservoir of gas in different phases (e.g. HI, CO).

Mergers are often claimed to be responsible for the triggering of the AGN in galaxies. However, the timescales of the radio activity are usually found to be much shorter ($\rm 10^6-10^7$ yr) than those computed for the merger events ($\rm 10^8-10^9$ yr), making the connection between the two phenomena weak (e.g. \citealp{emonts2006, struve2010,maccagni2014}). More compatible timescales ($\rm 10^5-10^7$ yr) come from multiple galaxy encounters that preceed the merger event and can increase the gas kinematical instabilities in the host galaxy, as suggested for explaining the origin of DDRGs \citep{schoenmakers2000}.

Alternatives that do not require any external contribution are instabilities in the accretion disk, for instance
caused by the radiation pressure induced warping \citep{pringle1997, czerny2009}, or chaotic accretion of cold gas clouds \citep{soker2009, gaspari2013, gaspari2017}. This last scenario looks particularly promising especially in gas rich galaxies like those analysed here. Indeed, local instabilities occurring in the observed gas structures (such as the HI disk in B2 0258+35) may cause an infall of clouds towards the galaxy centre, which may trigger the AGN frequently and intermittently (see \citealp{soker2009, gaspari2013, gaspari2017}).

\section{Conclusions and future work}

In this paper we have presented new multi-frequency observations of the source B2~0258+35 aimed at investigating the properties of the outer lobes initially identified by \cite{shulevski2012} at 1400 MHz. The lobes are further detected at 145 MHz with LOFAR and at 6600 MHz with SRT. The main findings are summarized below:
\medskip

(i) New observations with LOFAR at 145 MHz confirm the size ($\sim$265 kpc), relaxed shape and low surface brightness of the outer lobes previously observed at 1400 MHz by \cite{shulevski2012}. The emission from the outer lobes only represent $\rm \sim 6$\% of the entire source luminosity at 145 MHz. The extreme luminosity contrast between the inner and outer structure might be explained, on one hand, by strong adiabatic expansion of the plasma of the outer lobes in the ambient medium and, on the other hand, by a luminosity boost of the inner lobes due to compression of the magnetic field and the increased density of particles. 
\medskip

(ii) The S-shape morphology observed in the map at 1400 MHz is visible in the LOFAR map too, although much less pronounced due to the lower angular resolution. This structure may be interpreted as due to low power jets that connect the inner and outer lobes or to jet remnants from the previous cycle of activity. The S-shape may have originated from the jets bending due to a dense ISM or from precession.
\medskip

(iii) The combination of the new LOFAR and SRT observations with the data at 1400 MHz shows that the integrated spectrum of the outer northern lobe is not ultra steep and does not present a significant spectral curvature (SPC$\sim$0.2$\pm0.2$) with spectral indices equal to $\rm \alpha_{1400}^{145}=0.48\pm0.11$ and $\rm \alpha_{6600}^{1400}=0.69\pm0.20$. The spectral index in the outer southern lobe in the range 145-1400 MHz is $\rm \alpha_{1400}^{145}=0.73\pm0.07$.
\medskip 

(iv) By assuming a simple radiative model to describe the integrated spectrum of the outer northern lobe we have computed a first order upper limit to the radiative age equal to $\lesssim$110 Myr. This value is compatible with the dynamical age computed by \cite{shulevski2012} and provides an estimate of the jet quiescent time between the two subsequent phases of activity equal to $\lesssim$100 Myr. We stress that this value should be treated with care as it neglects mechanisms such as situ particle reacceleration, mixing and/or adiabatic compression that may alter the spectral shape. 
\medskip 

(v) Possible evolutionary scenarios to explain the source morphology and spectral properties of the outer lobes are the following: 1) the AGN has never switched off but the large-scale radio jets have only been temporarily disrupted. The outer lobes are still fuelled by fresh particles from the nucleus at a very low rate; 2) the active nucleus has switched off and on again after a short time so that the plasma in the outer lobes is still not aged; 3) the outer lobes are remnants of a previous AGN activity but in situ particle reacceleration, mixing and/or adiabatic compression prevent their spectrum from further steepening.
\medskip

(vi) As all the mechanisms mentioned above stop to be relevant on timescales longer than a few tens of Myr, it follows that the outer lobes are not very old remnants as the morphology alone would suggest and as it was previously claimed. Instead, we propose that either the outer lobes are still fuelled by the nuclear activity or the switch off must have happened only few tens of Myr ago, suggesting a jet quiescent phase shorter than a few tens of Myr.
\medskip

(vii) Future searches of restarted radio galaxies should bear in mind that not all low-surface brightness lobes and halos with amorphous shape that are found around compact, bright jets are necessarily ultra-steep spectrum remnants of past AGN activity. This is especially true for low power radio galaxies.

\bigskip

The morphological and the spectral characteristics of the source B2~0258+35 are quite exceptional and only a small number of other radio galaxies show some resemblance. Analysis of its polarization properties will be used to further investigate the nature of the outer lobes and the source evolution (Adebahr et al. in prep). 

To date, it is not clear whether these sources are intrinsically rare or whether they have just been missed by the old-generation instruments, due to their lack of sensitivity, especially at low frequency. LOFAR is the perfect instrument to clarify this as it provides at the same time high spatial resolution (20~arcsec up to 6~arcsec) and high sensitivity (with typical noise values up to 0.1 mJy $\rm beam^{-1}$ in 8 hours observations) at 145 MHz. By using LoTSS we are starting a systematic search for low surface brightness, extended emission around a sample of well-known compact radio galaxies ($<$ few kpc) that are located in the survey area. Low frequency data will be able to probe even the most aged particle population. With this we will get a better statistics of the presence of such extended structures around compact radio galaxies and we will be able to investigate their properties and nature in a statistical way.

\begin{acknowledgements}
The research leading to these results has received funding from the European Research Council under the European Union's Seventh Framework Programme (FP/2007-2013) / ERC Advanced Grant RADIOLIFE-320745. LOFAR, the Low Frequency Array designed and constructed by ASTRON (Netherlands Institute for Radio Astronomy), has facilities in several countries, that are owned by various parties (each with their own funding sources), and that are collectively operated by the International LOFAR Telescope (ILT) foundation under a joint scientific policy. The Sardinia Radio Telescope is funded by the Department of University and Research (MIUR), Italian Space Agency (ASI), and the Autonomous Region of Sardinia (RAS) and is operated as National Facility by the National Institute for Astrophysics (INAF). The development of the SARDARA back-end has been funded by the Autonomous Region of Sardinia (RAS) using resources from the Regional Law 7/2007 ”Promotion of the scientific research and technological innovation in Sardinia” in the context of the research project CRP-49231 (year 2011, PI Possenti): ”High resolution sampling of the Universe in the radio band: an unprecedented instrument to understand the fundamental laws of the nature”. We thank the staff of the GMRT that made these observations possible. GMRT is run by the National Centre for Radio Astrophysics of the Tata Institute of Fundamental Research. The National Radio Astronomy Observatory is a facility of the National Science Foundation operated under cooperative agreement by Associated Universities, Inc. This research has made use of the NASA/IPAC Extragalactic Database (NED), which is operated by the Jet Propulsion Laboratory, California Institute of Technology, under contract with the National Aeronautics and Space Administration. This research made use of APLpy, an open-source plotting package for Python hosted at http://aplpy.github.com. 
      
\end{acknowledgements}

\bibliographystyle{aa}
\bibliography{B20258+35_brienza.bib}

\begin{thebibliography}{135}
\expandafter\ifx\csname natexlab\endcsname\relax\def\natexlab#1{#1}\fi

\bibitem[{{Alexander}(1987)}]{alexandar1987}
{Alexander}, P. 1987, \mnras, 225, 27

\bibitem[{{Alvarez} {et~al.}(2000){Alvarez}, {Aparici}, {May}, \&
  {Reich}}]{alvarez2000}
{Alvarez}, H., {Aparici}, J., {May}, J., \& {Reich}, P. 2000, \aap, 355, 863

\bibitem[{{Baum} {et~al.}(1990){Baum}, {O'Dea}, {Murphy}, \& {de
  Bruyn}}]{baum1990}
{Baum}, S.~A., {O'Dea}, C.~P., {Murphy}, D.~W., \& {de Bruyn}, A.~G. 1990,
  \aap, 232, 19

\bibitem[{{Beck} \& {Krause}(2005)}]{beck2005}
{Beck}, R. \& {Krause}, M. 2005, Astronomische Nachrichten, 326, 414

\bibitem[{{Best} {et~al.}(2005){Best}, {Kauffmann}, {Heckman}, {Brinchmann},
  {Charlot}, {Ivezi{\'c}}, \& {White}}]{best2005}
{Best}, P.~N., {Kauffmann}, G., {Heckman}, T.~M., {et~al.} 2005, \mnras, 362,
  25

\bibitem[{{Bhatnagar} {et~al.}(2008){Bhatnagar}, {Cornwell}, {Golap}, \&
  {Uson}}]{bhatnagar2008}
{Bhatnagar}, S., {Cornwell}, T.~J., {Golap}, K., \& {Uson}, J.~M. 2008, \aap,
  487, 419

\bibitem[{{Bicknell} {et~al.}(2018){Bicknell}, {Mukherjee}, {Wagner},
  {Sutherland}, \& {Nesvadba}}]{bicknell2018}
{Bicknell}, G.~V., {Mukherjee}, D., {Wagner}, A.~Y., {Sutherland}, R.~S., \&
  {Nesvadba}, N.~P.~H. 2018, \mnras, 475, 3493

\bibitem[{{Bolli} {et~al.}(2015){Bolli}, {Orlati}, {Stringhetti}, {Orfei},
  {Righini}, {Ambrosini}, {Bartolini}, {Bortolotti}, {Buffa}, {Buttu},
  {Cattani}, {D'Amico}, {Deiana}, {Fara}, {Fiocchi}, {Gaudiomonte},
  {Maccaferri}, {Mariotti}, {Marongiu}, {Melis}, {Migoni}, {Morsiani}, {Nanni},
  {Nasyr}, {Pellizzoni}, {Pisanu}, {Poloni}, {Poppi}, {Porceddu}, {Prandoni},
  {Roda}, {Roma}, {Scalambra}, {Serra}, {Trois}, {Valente}, {Vargiu}, \&
  {Zacchiroli}}]{bolli2015}
{Bolli}, P., {Orlati}, A., {Stringhetti}, L., {et~al.} 2015, Journal of
  Astronomical Instrumentation, 4, 1550008

\bibitem[{{Bondi}(1952)}]{bondi1952}
{Bondi}, H. 1952, \mnras, 112, 195

\bibitem[{{Brienza} {et~al.}(2017){Brienza}, {Godfrey}, {Morganti}, {Prandoni},
  {Harwood}, {Mahony}, {Hardcastle}, {Murgia}, {R{\"o}ttgering}, {Shimwell}, \&
  {Shulevski}}]{brienza2017}
{Brienza}, M., {Godfrey}, L., {Morganti}, R., {et~al.} 2017, \aap, 606, A98

\bibitem[{{Brienza} {et~al.}(2016){Brienza}, {Godfrey}, {Morganti}, {Vilchez},
  {Maddox}, {Murgia}, {Orru}, {Shulevski}, {Best}, {Br{\"u}ggen}, {Harwood},
  {Jamrozy}, {Jarvis}, {Mahony}, {McKean}, \& {R{\"o}ttgering}}]{brienza2016}
{Brienza}, M., {Godfrey}, L., {Morganti}, R., {et~al.} 2016, \aap, 585, A29

\bibitem[{{Brown} {et~al.}(2011){Brown}, {Jannuzi}, {Floyd}, \&
  {Mould}}]{brown2011}
{Brown}, M.~J.~I., {Jannuzi}, B.~T., {Floyd}, D.~J.~E., \& {Mould}, J.~R. 2011,
  \apjl, 731, L41

\bibitem[{{Callingham} {et~al.}(2015){Callingham}, {Gaensler}, {Ekers},
  {Tingay}, {Wayth}, {Morgan}, {Bernardi}, {Bell}, {Bhat}, {Bowman}, {Briggs},
  {Cappallo}, {Deshpande}, {Ewall-Wice}, {Feng}, {Greenhill}, {Hazelton},
  {Hindson}, {Hurley-Walker}, {Jacobs}, {Johnston-Hollitt}, {Kaplan},
  {Kudrayvtseva}, {Lenc}, {Lonsdale}, {McKinley}, {McWhirter}, {Mitchell},
  {Morales}, {Morgan}, {Oberoi}, {Offringa}, {Ord}, {Pindor}, {Prabu},
  {Procopio}, {Riding}, {Srivani}, {Subrahmanyan}, {Udaya Shankar}, {Webster},
  {Williams}, \& {Williams}}]{callingham2015}
{Callingham}, J.~R., {Gaensler}, B.~M., {Ekers}, R.~D., {et~al.} 2015, \apj,
  809, 168

\bibitem[{{Cohen} {et~al.}(2007){Cohen}, {Lane}, {Cotton}, {Kassim}, {Lazio},
  {Perley}, {Condon}, \& {Erickson}}]{cohen2007}
{Cohen}, A.~S., {Lane}, W.~M., {Cotton}, W.~D., {et~al.} 2007, \aj, 134, 1245

\bibitem[{{Condon} {et~al.}(1998){Condon}, {Cotton}, {Greisen}, {Yin},
  {Perley}, {Taylor}, \& {Broderick}}]{condon1998}
{Condon}, J.~J., {Cotton}, W.~D., {Greisen}, E.~W., {et~al.} 1998, \aj, 115,
  1693

\bibitem[{{Cornwell} \& {Perley}(1992)}]{cornwell1992}
{Cornwell}, T.~J. \& {Perley}, R.~A. 1992, \aap, 261, 353

\bibitem[{{Croston}(2008)}]{croston2008}
{Croston}, J.~H. 2008, in Astronomical Society of the Pacific Conference
  Series, Vol. 386, Extragalactic Jets: Theory and Observation from Radio to
  Gamma Ray, ed. T.~A. {Rector} \& D.~S. {De Young}, 335

\bibitem[{{Czerny} {et~al.}(2009){Czerny}, {Siemiginowska}, {Janiuk},
  {Nikiel-Wroczy{\'n}ski}, \& {Stawarz}}]{czerny2009}
{Czerny}, B., {Siemiginowska}, A., {Janiuk}, A., {Nikiel-Wroczy{\'n}ski}, B.,
  \& {Stawarz}, {\L}. 2009, \apj, 698, 840

\bibitem[{{de Gasperin} {et~al.}(2012){de Gasperin}, {Orr{\'u}}, {Murgia},
  {Merloni}, {Falcke}, {Beck}, {Beswick}, {B{\^i}rzan}, {Bonafede},
  {Br{\"u}ggen}, {Brunetti}, {Chy{\.z}y}, {Conway}, {Croston}, {En{\ss}lin},
  {Ferrari}, {Heald}, {Heidenreich}, {Jackson}, {Macario}, {McKean}, {Miley},
  {Morganti}, {Offringa}, {Pizzo}, {Rafferty}, {R{\"o}ttgering}, {Shulevski},
  {Steinmetz}, {Tasse}, {van der Tol}, {van Driel}, {van Weeren}, {van
  Zwieten}, {Alexov}, {Anderson}, {Asgekar}, {Avruch}, {Bell}, {Bell},
  {Bentum}, {Bernardi}, {Best}, {Breitling}, {Broderick}, {Butcher}, {Ciardi},
  {Dettmar}, {Eisloeffel}, {Frieswijk}, {Gankema}, {Garrett}, {Gerbers},
  {Griessmeier}, {Gunst}, {Hassall}, {Hessels}, {Hoeft}, {Horneffer},
  {Karastergiou}, {K{\"o}hler}, {Koopman}, {Kuniyoshi}, {Kuper}, {Maat},
  {Mann}, {Mevius}, {Mulcahy}, {Munk}, {Nijboer}, {Noordam}, {Paas}, {Pandey},
  {Pandey}, {Polatidis}, {Reich}, {Schoenmakers}, {Sluman}, {Smirnov}, {Sobey},
  {Stappers}, {Swinbank}, {Tagger}, {Tang}, {van Bemmel}, {van Cappellen}, {van
  Duin}, {van Haarlem}, {van Leeuwen}, {Vermeulen}, {Vocks}, {White}, {Wise},
  {Wucknitz}, \& {Zarka}}]{degasperin2012}
{de Gasperin}, F., {Orr{\'u}}, E., {Murgia}, M., {et~al.} 2012, \aap, 547, A56

\bibitem[{{Di Matteo} {et~al.}(2005){Di Matteo}, {Springel}, \&
  {Hernquist}}]{dimatteo2005}
{Di Matteo}, T., {Springel}, V., \& {Hernquist}, L. 2005, \nat, 433, 604

\bibitem[{{Douglas} {et~al.}(1996){Douglas}, {Bash}, {Bozyan}, {Torrence}, \&
  {Wolfe}}]{douglas1996}
{Douglas}, J.~N., {Bash}, F.~N., {Bozyan}, F.~A., {Torrence}, G.~W., \&
  {Wolfe}, C. 1996, \aj, 111, 1945

\bibitem[{{Dressel} \& {Condon}(1978)}]{dressel1978}
{Dressel}, L.~L. \& {Condon}, J.~J. 1978, \apjs, 36, 53

\bibitem[{{Eilek}(2014)}]{eilek2014}
{Eilek}, J.~A. 2014, New Journal of Physics, 16, 045001

\bibitem[{{Emonts}(2006)}]{emonts2006}
{Emonts}, B.~H.~C. 2006, PhD thesis, University of Groningen

\bibitem[{{Emonts} {et~al.}(2010){Emonts}, {Morganti}, {Struve}, {Oosterloo},
  {van Moorsel}, {Tadhunter}, {van der Hulst}, {Brogt}, {Holt}, \&
  {Mirabal}}]{emonts2010}
{Emonts}, B.~H.~C., {Morganti}, R., {Struve}, C., {et~al.} 2010, \mnras, 406,
  987

\bibitem[{{Fabian}(2012)}]{fabian2012}
{Fabian}, A.~C. 2012, \araa, 50, 455

\bibitem[{{Falceta-Gon{\c c}alves} {et~al.}(2010){Falceta-Gon{\c c}alves},
  {Caproni}, {Abraham}, {Teixeira}, \& {de Gouveia Dal Pino}}]{falceta2010}
{Falceta-Gon{\c c}alves}, D., {Caproni}, A., {Abraham}, Z., {Teixeira}, D.~M.,
  \& {de Gouveia Dal Pino}, E.~M. 2010, \apjl, 713, L74

\bibitem[{{Fanti}(2009)}]{fanti2009}
{Fanti}, C. 2009, Astronomische Nachrichten, 330, 120

\bibitem[{{Frank} {et~al.}(2016){Frank}, {Morganti}, {Oosterloo}, {Nyland}, \&
  {Serra}}]{frank2016}
{Frank}, B.~S., {Morganti}, R., {Oosterloo}, T., {Nyland}, K., \& {Serra}, P.
  2016, \aap, 592, A94

\bibitem[{{Gaspari} {et~al.}(2012){Gaspari}, {Brighenti}, \&
  {Temi}}]{gaspari2012}
{Gaspari}, M., {Brighenti}, F., \& {Temi}, P. 2012, \mnras, 424, 190

\bibitem[{{Gaspari} {et~al.}(2013){Gaspari}, {Ruszkowski}, \&
  {Oh}}]{gaspari2013}
{Gaspari}, M., {Ruszkowski}, M., \& {Oh}, S.~P. 2013, \mnras, 432, 3401

\bibitem[{{Gaspari} {et~al.}(2017){Gaspari}, {Temi}, \&
  {Brighenti}}]{gaspari2017}
{Gaspari}, M., {Temi}, P., \& {Brighenti}, F. 2017, \mnras, 466, 677

\bibitem[{{Giovannini} {et~al.}(2001){Giovannini}, {Cotton}, {Feretti}, {Lara},
  \& {Venturi}}]{giovannini2001}
{Giovannini}, G., {Cotton}, W.~D., {Feretti}, L., {Lara}, L., \& {Venturi}, T.
  2001, \apj, 552, 508

\bibitem[{{Giroletti} {et~al.}(2005){Giroletti}, {Giovannini}, \&
  {Taylor}}]{giroletti2005}
{Giroletti}, M., {Giovannini}, G., \& {Taylor}, G.~B. 2005, \aap, 441, 89

\bibitem[{{Godfrey} {et~al.}(2017){Godfrey}, {Morganti}, \&
  {Brienza}}]{godfrey2017}
{Godfrey}, L.~E.~H., {Morganti}, R., \& {Brienza}, M. 2017, \mnras, 471, 891

\bibitem[{{Gong} {et~al.}(2011){Gong}, {Li}, \& {Zhang}}]{gong2011}
{Gong}, B.~P., {Li}, Y.~P., \& {Zhang}, H.~C. 2011, \apjl, 734, L32

\bibitem[{{Govoni} {et~al.}(2017){Govoni}, {Murgia}, {Vacca}, {Loi}, {Girardi},
  {Gastaldello}, {Giovannini}, {Feretti}, {Paladino}, {Carretti}, {Concu},
  {Melis}, {Poppi}, {Valente}, {Bernardi}, {Bonafede}, {Boschin}, {Brienza},
  {Clarke}, {Colafrancesco}, {de Gasperin}, {Eckert}, {En{\ss}lin}, {Ferrari},
  {Gregorini}, {Johnston-Hollitt}, {Junklewitz}, {Orr{\`u}}, {Parma}, {Perley},
  {Rossetti}, {B Taylor}, \& {Vazza}}]{govoni2017}
{Govoni}, F., {Murgia}, M., {Vacca}, V., {et~al.} 2017, \aap, 603, A122

\bibitem[{{Gregory} \& {Condon}(1991)}]{gregory1991}
{Gregory}, P.~C. \& {Condon}, J.~J. 1991, \apjs, 75, 1011

\bibitem[{{Griffith} {et~al.}(1990){Griffith}, {Langston}, {Heflin}, {Conner},
  {Lehar}, \& {Burke}}]{griffith1990}
{Griffith}, M., {Langston}, G., {Heflin}, M., {et~al.} 1990, \apjs, 74, 129

\bibitem[{{Hales} {et~al.}(1993){Hales}, {Baldwin}, \& {Warner}}]{hales1993}
{Hales}, S.~E.~G., {Baldwin}, J.~E., \& {Warner}, P.~J. 1993, \mnras, 263, 25

\bibitem[{{Hardcastle} {et~al.}(2009){Hardcastle}, {Cheung}, {Feain}, \&
  {Stawarz}}]{hardcastle2009}
{Hardcastle}, M.~J., {Cheung}, C.~C., {Feain}, I.~J., \& {Stawarz}, {\L}. 2009,
  \mnras, 393, 1041

\bibitem[{{Hardcastle} {et~al.}(2007){Hardcastle}, {Evans}, \&
  {Croston}}]{hardcastle2007}
{Hardcastle}, M.~J., {Evans}, D.~A., \& {Croston}, J.~H. 2007, \mnras, 376,
  1849

\bibitem[{{Heald} {et~al.}(2010){Heald}, {McKean}, {Pizzo}, {van Diepen}, {van
  Zwieten}, {van Weeren}, {Rafferty}, {van der Tol}, {Birzan}, {Shulevski},
  {Swinbank}, {Orru}, {de Gasperin}, {Ker}, {Bonafede}, {Macario}, \&
  {Ferrari}}]{heald2010}
{Heald}, G., {McKean}, J., {Pizzo}, R., {et~al.} 2010, in ISKAF2010 Science
  Meeting, 57

\bibitem[{{Ho} {et~al.}(1997){Ho}, {Filippenko}, \& {Sargent}}]{ho1997}
{Ho}, L.~C., {Filippenko}, A.~V., \& {Sargent}, W.~L.~W. 1997, \apjs, 112, 315

\bibitem[{{Hurley-Walker} {et~al.}(2017){Hurley-Walker}, {Callingham},
  {Hancock}, {Franzen}, {Hindson}, {Kapi{\'n}ska}, {Morgan}, {Offringa},
  {Wayth}, {Wu}, {Zheng}, {Murphy}, {Bell}, {Dwarakanath}, {For}, {Gaensler},
  {Johnston-Hollitt}, {Lenc}, {Procopio}, {Staveley-Smith}, {Ekers}, {Bowman},
  {Briggs}, {Cappallo}, {Deshpande}, {Greenhill}, {Hazelton}, {Kaplan},
  {Lonsdale}, {McWhirter}, {Mitchell}, {Morales}, {Morgan}, {Oberoi}, {Ord},
  {Prabu}, {Shankar}, {Srivani}, {Subrahmanyan}, {Tingay}, {Webster},
  {Williams}, \& {Williams}}]{hurleywalker2017}
{Hurley-Walker}, N., {Callingham}, J.~R., {Hancock}, P.~J., {et~al.} 2017,
  \mnras, 464, 1146

\bibitem[{{Hurley-Walker} {et~al.}(2015){Hurley-Walker}, {Johnston-Hollitt},
  {Ekers}, {Hunstead}, {Sadler}, {Hindson}, {Hancock}, {Bernardi}, {Bowman},
  {Briggs}, {Cappallo}, {Corey}, {Deshpande}, {Emrich}, {Gaensler}, {Goeke},
  {Greenhill}, {Hazelton}, {Hewitt}, {Kaplan}, {Kasper}, {Kratzenberg},
  {Lonsdale}, {Lynch}, {Mitchell}, {McWhirter}, {Morales}, {Morgan}, {Oberoi},
  {Offringa}, {Ord}, {Prabu}, {Rogers}, {Roshi}, {Shankar}, {Srivani},
  {Subrahmanyan}, {Tingay}, {Waterson}, {Wayth}, {Webster}, {Whitney},
  {Williams}, \& {Williams}}]{hurleywalker2015}
{Hurley-Walker}, N., {Johnston-Hollitt}, M., {Ekers}, R., {et~al.} 2015,
  \mnras, 447, 2468

\bibitem[{{Intema}(2014)}]{intema2014}
{Intema}, H.~T. 2014, in Astronomical Society of India Conference Series,
  Vol.~13, Astronomical Society of India Conference Series

\bibitem[{{Intema} {et~al.}(2017){Intema}, {Jagannathan}, {Mooley}, \&
  {Frail}}]{intema2017}
{Intema}, H.~T., {Jagannathan}, P., {Mooley}, K.~P., \& {Frail}, D.~A. 2017,
  \aap, 598, A78

\bibitem[{{Jamrozy} {et~al.}(2007){Jamrozy}, {Konar}, {Saikia}, {Stawarz},
  {Mack}, \& {Siemiginowska}}]{jamrozy2007}
{Jamrozy}, M., {Konar}, C., {Saikia}, D.~J., {et~al.} 2007, \mnras, 378, 581

\bibitem[{{Jamrozy} {et~al.}(2009){Jamrozy}, {Saikia}, \&
  {Konar}}]{jamrozy2009}
{Jamrozy}, M., {Saikia}, D.~J., \& {Konar}, C. 2009, \mnras, 399, L141

\bibitem[{{Joshi} {et~al.}(2011){Joshi}, {Nandi}, {Saikia}, {Ishwara-Chandra},
  \& {Konar}}]{joshi2012}
{Joshi}, S.~A., {Nandi}, S., {Saikia}, D.~J., {Ishwara-Chandra}, C.~H., \&
  {Konar}, C. 2011, \mnras, 414, 1397

\bibitem[{{Kaiser} {et~al.}(2000){Kaiser}, {Schoenmakers}, \&
  {R{\"o}ttgering}}]{kaiser2000}
{Kaiser}, C.~R., {Schoenmakers}, A.~P., \& {R{\"o}ttgering}, H.~J.~A. 2000,
  \mnras, 315, 381

\bibitem[{{Kardashev}(1962)}]{kardashev1962}
{Kardashev}, N.~S. 1962, \sovast, 6, 317

\bibitem[{{Kellermann}(1964)}]{kellerman1964}
{Kellermann}, K.~I. 1964, \apj, 140, 969

\bibitem[{{Kellermann} \& {Pauliny-Toth}(1973)}]{kellerman1973}
{Kellermann}, K.~I. \& {Pauliny-Toth}, I.~I.~K. 1973, \aj, 78, 828

\bibitem[{{Kharb} {et~al.}(2006){Kharb}, {O'Dea}, {Baum}, {Colbert}, \&
  {Xu}}]{kharb2006}
{Kharb}, P., {O'Dea}, C.~P., {Baum}, S.~A., {Colbert}, E.~J.~M., \& {Xu}, C.
  2006, \apj, 652, 177

\bibitem[{{Klein} {et~al.}(2003){Klein}, {Mack}, {Gregorini}, \&
  {Vigotti}}]{klein2003}
{Klein}, U., {Mack}, K.-H., {Gregorini}, L., \& {Vigotti}, M. 2003, \aap, 406,
  579

\bibitem[{{Konar} \& {Hardcastle}(2013)}]{konar2013a}
{Konar}, C. \& {Hardcastle}, M.~J. 2013, \mnras, 436, 1595

\bibitem[{{Konar} {et~al.}(2013){Konar}, {Hardcastle}, {Jamrozy}, \&
  {Croston}}]{konar2013b}
{Konar}, C., {Hardcastle}, M.~J., {Jamrozy}, M., \& {Croston}, J.~H. 2013,
  \mnras, 430, 2137

\bibitem[{{Ku{\'z}micz} {et~al.}(2017){Ku{\'z}micz}, {Jamrozy},
  {Kozie{\l}-Wierzbowska}, \& {We{\.z}gowiec}}]{kuzmicz2017}
{Ku{\'z}micz}, A., {Jamrozy}, M., {Kozie{\l}-Wierzbowska}, D., \&
  {We{\.z}gowiec}, M. 2017, \mnras, 471, 3806

\bibitem[{{Laing}(1980)}]{laing1980}
{Laing}, R.~A. 1980, \mnras, 193, 439

\bibitem[{{Loi} {et~al.}(2017){Loi}, {Murgia}, {Govoni}, {Vacca}, {Feretti},
  {Giovannini}, {Carretti}, {Gastaldello}, {Girardi}, {Vazza}, {Concu},
  {Melis}, {Paladino}, {Poppi}, {Valente}, {Boschin}, {Clarke},
  {Colafrancesco}, {En{\ss}lin}, {Ferrari}, {de Gasperin}, {Gregorini},
  {Johnston-Hollitt}, {Junklewitz}, {Orr{\`u}}, {Parma}, {Perley}, \&
  {Taylor}}]{loi2017}
{Loi}, F., {Murgia}, M., {Govoni}, F., {et~al.} 2017, \mnras, 472, 3605

\bibitem[{{Lu} \& {Zhou}(2005)}]{lu2005}
{Lu}, J.-F. \& {Zhou}, B.-Y. 2005, \apjl, 635, L17

\bibitem[{{Luo} {et~al.}(2007){Luo}, {Yang}, {Cui}, {Liu}, \& {Shen}}]{luo2007}
{Luo}, W.-F., {Yang}, J., {Cui}, L., {Liu}, X., \& {Shen}, Z.-Q. 2007, \cjaa,
  7, 611

\bibitem[{{Maccagni} {et~al.}(2014){Maccagni}, {Morganti}, {Oosterloo}, \&
  {Mahony}}]{maccagni2014}
{Maccagni}, F.~M., {Morganti}, R., {Oosterloo}, T.~A., \& {Mahony}, E.~K. 2014,
  \aap, 571, A67

\bibitem[{{Machalski} {et~al.}(2016){Machalski}, {Jamrozy}, {Stawarz}, \&
  {We{\.z}gowiec}}]{machalski2016}
{Machalski}, J., {Jamrozy}, M., {Stawarz}, {\L}., \& {We{\.z}gowiec}, M. 2016,
  \aap, 595, A46

\bibitem[{{Mahatma} {et~al.}(2018){Mahatma}, {Hardcastle}, {Williams},
  {Brienza}, {Br{\"u}ggen}, {Croston}, {Gurkan}, {Harwood},
  {Kunert-Bajraszewska}, {Morganti}, {R{\"o}ttgering}, {Shimwell}, \&
  {Tasse}}]{mahatma2018}
{Mahatma}, V.~H., {Hardcastle}, M.~J., {Williams}, W.~L., {et~al.} 2018,
  \mnras, 475, 4557

\bibitem[{{Mahony} {et~al.}(2016){Mahony}, {Morganti}, {Prandoni}, {van
  Bemmel}, {Shimwell}, {Brienza}, {Best}, {Br{\"u}ggen}, {Calistro Rivera}, {de
  Gasperin}, {Hardcastle}, {Harwood}, {Heald}, {Jarvis}, {Mandal}, {Miley},
  {Retana-Montenegro}, {R{\"o}ttgering}, {Sabater}, {Tasse}, {van Velzen}, {van
  Weeren}, {Williams}, \& {White}}]{mahony2016}
{Mahony}, E.~K., {Morganti}, R., {Prandoni}, I., {et~al.} 2016, \mnras, 463,
  2997

\bibitem[{{Massaglia} {et~al.}(2016){Massaglia}, {Bodo}, {Rossi}, {Capetti}, \&
  {Mignone}}]{massaglia2016}
{Massaglia}, S., {Bodo}, G., {Rossi}, P., {Capetti}, S., \& {Mignone}, A. 2016,
  \aap, 596, A12

\bibitem[{{McKinley} {et~al.}(2013){McKinley}, {Briggs}, {Gaensler}, {Feain},
  {Bernardi}, {Wayth}, {Johnston-Hollitt}, {Offringa}, {Arcus}, {Barnes},
  {Bowman}, {Bunton}, {Cappallo}, {Corey}, {Deshpande}, {deSouza}, {Emrich},
  {Goeke}, {Greenhill}, {Hazelton}, {Herne}, {Hewitt}, {Kaplan}, {Kasper},
  {Kincaid}, {Koenig}, {Kratzenberg}, {Lonsdale}, {Lynch}, {McWhirter},
  {Mitchell}, {Morales}, {Morgan}, {Oberoi}, {Ord}, {Pathikulangara}, {Prabu},
  {Remillard}, {Rogers}, {Roshi}, {Salah}, {Sault}, {Shankar}, {Srivani},
  {Stevens}, {Subrahmanyan}, {Tingay}, {Waterson}, {Webster}, {Whitney},
  {Williams}, {Williams}, \& {Wyithe}}]{mckinley2013}
{McKinley}, B., {Briggs}, F., {Gaensler}, B.~M., {et~al.} 2013, \mnras, 436,
  1286

\bibitem[{{McKinley} {et~al.}(2018){McKinley}, {Tingay}, {Carretti}, {Ellis},
  {Bland-Hawthorn}, {Morganti}, {Line}, {McDonald}, {Veilleux}, {Wahl Olsen},
  {Sidonio}, {Ekers}, {Offringa}, {Procopio}, {Pindor}, {Wayth},
  {Hurley-Walker}, {Bernardi}, {Gaensler}, {Haverkorn}, {Kesteven}, {Poppi}, \&
  {Staveley-Smith}}]{mckinley2017}
{McKinley}, B., {Tingay}, S.~J., {Carretti}, E., {et~al.} 2018, \mnras, 474,
  4056

\bibitem[{{McMullin} {et~al.}(2007){McMullin}, {Waters}, {Schiebel}, {Young},
  \& {Golap}}]{mcmullin2007}
{McMullin}, J.~P., {Waters}, B., {Schiebel}, D., {Young}, W., \& {Golap}, K.
  2007, in Astronomical Society of the Pacific Conference Series, Vol. 376,
  Astronomical Data Analysis Software and Systems XVI, ed. R.~A. {Shaw},
  F.~{Hill}, \& D.~J. {Bell}, 127

\bibitem[{{McNamara} \& {Nulsen}(2007)}]{mcnamara2007}
{McNamara}, B.~R. \& {Nulsen}, P.~E.~J. 2007, \araa, 45, 117

\bibitem[{{Melis} {et~al.}(2018){Melis}, {Concu}, {Trois}, {Possenti},
  {Bocchinu}, {Bolli}, {Burgay}, {Carretti}, {Castangia}, {Casu}, {Pestellini},
  {Corongiu}, {D'Amico}, {Egron}, {Govoni}, {Iacolina}, {Murgia}, {Pellizzoni},
  {Perrodin}, {Pilia}, {Pisanu}, {Poddighe}, {Poppi}, {Porceddu}, {Tarchi},
  {Vacca}, {Aresu}, {Bachetti}, {Barbaro}, {Casula}, {Ladu}, {Leurini}, {Loi},
  {Loru}, {Marongiu}, {Maxia}, {Mazzarella}, {Migoni}, {Montisci}, {Valente},
  \& {Vargiu}}]{melis2018}
{Melis}, A., {Concu}, R., {Trois}, A., {et~al.} 2018, Journal of Astronomical
  Instrumentation, 7, 1850004

\bibitem[{{Morganti}(2017)}]{morganti2017}
{Morganti}, R. 2017, Nature Astronomy, 1, 596

\bibitem[{{Morganti} {et~al.}(2011){Morganti}, {Holt}, {Tadhunter}, {Ramos
  Almeida}, {Dicken}, {Inskip}, {Oosterloo}, \& {Tzioumis}}]{morganti2011}
{Morganti}, R., {Holt}, J., {Tadhunter}, C., {et~al.} 2011, \aap, 535, A97

\bibitem[{{Morganti} {et~al.}(1999){Morganti}, {Killeen}, {Ekers}, \&
  {Oosterloo}}]{morganti1999}
{Morganti}, R., {Killeen}, N.~E.~B., {Ekers}, R.~D., \& {Oosterloo}, T.~A.
  1999, \mnras, 307, 750

\bibitem[{{Murgia}(2003)}]{murgia2003}
{Murgia}, M. 2003, \pasa, 20, 19

\bibitem[{{Murgia} {et~al.}(1999){Murgia}, {Fanti}, {Fanti}, {Gregorini},
  {Klein}, {Mack}, \& {Vigotti}}]{murgia1999}
{Murgia}, M., {Fanti}, C., {Fanti}, R., {et~al.} 1999, \aap, 345, 769

\bibitem[{{Murgia} {et~al.}(2016){Murgia}, {Govoni}, {Carretti}, {Melis},
  {Concu}, {Trois}, {Loi}, {Vacca}, {Tarchi}, {Castangia}, {Possenti},
  {Bocchinu}, {Burgay}, {Casu}, {Pellizzoni}, {Pisanu}, {Poddighe}, {Poppi},
  {D'Amico}, {Bachetti}, {Corongiu}, {Egron}, {Iacolina}, {Ladu}, {Marongiu},
  {Migoni}, {Perrodin}, {Pilia}, {Valente}, \& {Vargiu}}]{murgia2016}
{Murgia}, M., {Govoni}, F., {Carretti}, E., {et~al.} 2016, \mnras, 461, 3516

\bibitem[{{Murgia} {et~al.}(2011){Murgia}, {Parma}, {Mack}, {de Ruiter},
  {Fanti}, {Govoni}, {Tarchi}, {Giacintucci}, \& {Markevitch}}]{murgia2011}
{Murgia}, M., {Parma}, P., {Mack}, K.-H., {et~al.} 2011, \aap, 526, A148

\bibitem[{{Nawaz} {et~al.}(2016){Nawaz}, {Bicknell}, {Wagner}, {Sutherland}, \&
  {McNamara}}]{nawaz2016}
{Nawaz}, M.~A., {Bicknell}, G.~V., {Wagner}, A.~Y., {Sutherland}, R.~S., \&
  {McNamara}, B.~R. 2016, \mnras, 458, 802

\bibitem[{{Nixon} \& {King}(2013)}]{nixon2013}
{Nixon}, C. \& {King}, A. 2013, \apjl, 765, L7

\bibitem[{{Noordermeer} {et~al.}(2005){Noordermeer}, {van der Hulst},
  {Sancisi}, {Swaters}, \& {van Albada}}]{noordermeer2005}
{Noordermeer}, E., {van der Hulst}, J.~M., {Sancisi}, R., {Swaters}, R.~A., \&
  {van Albada}, T.~S. 2005, \aap, 442, 137

\bibitem[{{O'Dea} \& {Baum}(1997)}]{odea1997}
{O'Dea}, C.~P. \& {Baum}, S.~A. 1997, \aj, 113, 148

\bibitem[{{Orienti} \& {Dallacasa}(2008)}]{orienti2008}
{Orienti}, M. \& {Dallacasa}, D. 2008, \aap, 487, 885

\bibitem[{{Orienti} \& {Dallacasa}(2014)}]{orienti2014}
{Orienti}, M. \& {Dallacasa}, D. 2014, \mnras, 438, 463

\bibitem[{{Orr{\`u}} {et~al.}(2015){Orr{\`u}}, {van Velzen}, {Pizzo},
  {Yatawatta}, {Paladino}, {Iacobelli}, {Murgia}, {Falcke}, {Morganti}, {de
  Bruyn}, {Ferrari}, {Anderson}, {Bonafede}, {Mulcahy}, {Asgekar}, {Avruch},
  {Beck}, {Bell}, {van Bemmel}, {Bentum}, {Bernardi}, {Best}, {Breitling},
  {Broderick}, {Br{\"u}ggen}, {Butcher}, {Ciardi}, {Conway}, {Corstanje}, {de
  Geus}, {Deller}, {Duscha}, {Eisl{\"o}ffel}, {Engels}, {Frieswijk}, {Garrett},
  {Grie{\ss}meier}, {Gunst}, {Hamaker}, {Heald}, {Hoeft}, {van der Horst},
  {Intema}, {Juette}, {Kohler}, {Kondratiev}, {Kuniyoshi}, {Kuper}, {Loose},
  {Maat}, {Mann}, {Markoff}, {McFadden}, {McKay-Bukowski}, {Miley}, {Moldon},
  {Molenaar}, {Munk}, {Nelles}, {Paas}, {Pandey-Pommier}, {Pandey}, {Pietka},
  {Polatidis}, {Reich}, {R{\"o}ttgering}, {Rowlinson}, {Scaife},
  {Schoenmakers}, {Schwarz}, {Serylak}, {Shulevski}, {Smirnov}, {Steinmetz},
  {Stewart}, {Swinbank}, {Tagger}, {Tasse}, {Thoudam}, {Toribio}, {Vermeulen},
  {Vocks}, {van Weeren}, {Wijers}, {Wise}, \& {Wucknitz}}]{orru2015}
{Orr{\`u}}, E., {van Velzen}, S., {Pizzo}, R.~F., {et~al.} 2015, \aap, 584,
  A112

\bibitem[{{Owen} {et~al.}(2000){Owen}, {Eilek}, \& {Kassim}}]{owen2000}
{Owen}, F.~N., {Eilek}, J.~A., \& {Kassim}, N.~E. 2000, \apj, 543, 611

\bibitem[{{Pacholczyk}(1970)}]{pacholczyc1970}
{Pacholczyk}, A.~G. 1970, {Radio astrophysics. Nonthermal processes in galactic
  and extragalactic sources}

\bibitem[{{Parma} {et~al.}(2007){Parma}, {Murgia}, {de Ruiter}, {Fanti},
  {Mack}, \& {Govoni}}]{parma2007}
{Parma}, P., {Murgia}, M., {de Ruiter}, H.~R., {et~al.} 2007, \aap, 470, 875

\bibitem[{{Perley} \& {Butler}(2013)}]{perley2013}
{Perley}, R.~A. \& {Butler}, B.~J. 2013, \apjs, 204, 19

\bibitem[{{Pilkington} \& {Scott}(1965)}]{pilkington1965}
{Pilkington}, J.~D.~H. \& {Scott}, J.~F. 1965, \memras, 69, 183

\bibitem[{{Pizzolato} \& {Soker}(2010)}]{pizzolato2010}
{Pizzolato}, F. \& {Soker}, N. 2010, \mnras, 408, 961

\bibitem[{{Prandoni} {et~al.}(2007){Prandoni}, {Laing}, {Parma}, {de Ruiter},
  {Montenegro-Montes}, \& {Wilson}}]{prandoni2007}
{Prandoni}, I., {Laing}, R.~A., {Parma}, P., {et~al.} 2007, in Astronomical
  Society of the Pacific Conference Series, Vol. 375, From Z-Machines to ALMA:
  (Sub)Millimeter Spectroscopy of Galaxies, ed. A.~J. {Baker}, J.~{Glenn},
  A.~I. {Harris}, J.~G. {Mangum}, \& M.~S. {Yun}, 271

\bibitem[{{Prandoni} {et~al.}(2017){Prandoni}, {Murgia}, {Tarchi}, {Burgay},
  {Castangia}, {Egron}, {Govoni}, {Pellizzoni}, {Ricci}, {Righini},
  {Bartolini}, {Casu}, {Corongiu}, {Iacolina}, {Melis}, {Nasir}, {Orlati},
  {Perrodin}, {Poppi}, {Trois}, {Vacca}, {Zanichelli}, {Bachetti}, {Buttu},
  {Comoretto}, {Concu}, {Fara}, {Gaudiomonte}, {Loi}, {Migoni}, {Orfei},
  {Pilia}, {Bolli}, {Carretti}, {D'Amico}, {Guidetti}, {Loru}, {Massi},
  {Pisanu}, {Porceddu}, {Ridolfi}, {Serra}, {Stanghellini}, {Tiburzi},
  {Tingay}, \& {Valente}}]{prandoni2017}
{Prandoni}, I., {Murgia}, M., {Tarchi}, A., {et~al.} 2017, \aap, 608, A40

\bibitem[{{Pringle}(1997)}]{pringle1997}
{Pringle}, J.~E. 1997, \mnras, 292, 136

\bibitem[{{Rengelink} {et~al.}(1997){Rengelink}, {Tang}, {de Bruyn}, {Miley},
  {Bremer}, {Roettgering}, \& {Bremer}}]{rengelink1997}
{Rengelink}, R.~B., {Tang}, Y., {de Bruyn}, A.~G., {et~al.} 1997, \aaps, 124,
  259

\bibitem[{{Saikia} \& {Jamrozy}(2009)}]{saikia2009}
{Saikia}, D.~J. \& {Jamrozy}, M. 2009, Bulletin of the Astronomical Society of
  India, 37, 63

\bibitem[{{Sanghera} {et~al.}(1995){Sanghera}, {Saikia}, {Luedke}, {Spencer},
  {Foulsham}, {Akujor}, \& {Tzioumis}}]{sanghera1995}
{Sanghera}, H.~S., {Saikia}, D.~J., {Luedke}, E., {et~al.} 1995, \aap, 295, 629

\bibitem[{{Saripalli} {et~al.}(2012){Saripalli}, {Subrahmanyan}, {Thorat},
  {Ekers}, {Hunstead}, {Johnston}, \& {Sadler}}]{saripalli2012}
{Saripalli}, L., {Subrahmanyan}, R., {Thorat}, K., {et~al.} 2012, \apjs, 199,
  27

\bibitem[{{Saxton} {et~al.}(2001){Saxton}, {Sutherland}, \&
  {Bicknell}}]{saxton2001}
{Saxton}, C.~J., {Sutherland}, R.~S., \& {Bicknell}, G.~V. 2001, \apj, 563, 103

\bibitem[{{Scaife} \& {Heald}(2012)}]{scaife2012}
{Scaife}, A.~M.~M. \& {Heald}, G.~H. 2012, \mnras, 423, L30

\bibitem[{{Schaye} {et~al.}(2015){Schaye}, {Crain}, {Bower}, {Furlong},
  {Schaller}, {Theuns}, {Dalla Vecchia}, {Frenk}, {McCarthy}, {Helly},
  {Jenkins}, {Rosas-Guevara}, {White}, {Baes}, {Booth}, {Camps}, {Navarro},
  {Qu}, {Rahmati}, {Sawala}, {Thomas}, \& {Trayford}}]{schaye2015}
{Schaye}, J., {Crain}, R.~A., {Bower}, R.~G., {et~al.} 2015, \mnras, 446, 521

\bibitem[{{Scheuer} \& {Williams}(1968)}]{scheuer1968}
{Scheuer}, P.~A.~G. \& {Williams}, P.~J.~S. 1968, \araa, 6, 321

\bibitem[{{Schoenmakers} {et~al.}(2000){Schoenmakers}, {de Bruyn},
  {R{\"o}ttgering}, {van der Laan}, \& {Kaiser}}]{schoenmakers2000}
{Schoenmakers}, A.~P., {de Bruyn}, A.~G., {R{\"o}ttgering}, H.~J.~A., {van der
  Laan}, H., \& {Kaiser}, C.~R. 2000, \mnras, 315, 371

\bibitem[{{Shabala} {et~al.}(2008){Shabala}, {Ash}, {Alexander}, \&
  {Riley}}]{shabala2008}
{Shabala}, S.~S., {Ash}, S., {Alexander}, P., \& {Riley}, J.~M. 2008, \mnras,
  388, 625

\bibitem[{{Shimwell} {et~al.}(2017){Shimwell}, {R{\"o}ttgering}, {Best},
  {Williams}, {Dijkema}, {de Gasperin}, {Hardcastle}, {Heald}, {Hoang},
  {Horneffer}, {Intema}, {Mahony}, {Mandal}, {Mechev}, {Morabito}, {Oonk},
  {Rafferty}, {Retana-Montenegro}, {Sabater}, {Tasse}, {van Weeren},
  {Br{\"u}ggen}, {Brunetti}, {Chy{\.z}y}, {Conway}, {Haverkorn}, {Jackson},
  {Jarvis}, {McKean}, {Miley}, {Morganti}, {White}, {Wise}, {van Bemmel},
  {Beck}, {Brienza}, {Bonafede}, {Calistro Rivera}, {Cassano}, {Clarke},
  {Cseh}, {Deller}, {Drabent}, {van Driel}, {Engels}, {Falcke}, {Ferrari},
  {Fr{\"o}hlich}, {Garrett}, {Harwood}, {Heesen}, {Hoeft}, {Horellou},
  {Israel}, {Kapi{\'n}ska}, {Kunert-Bajraszewska}, {McKay}, {Mohan},
  {Orr{\'u}}, {Pizzo}, {Prandoni}, {Schwarz}, {Shulevski}, {Sipior}, {Smith},
  {Sridhar}, {Steinmetz}, {Stroe}, {Varenius}, {van der Werf}, {Zensus}, \&
  {Zwart}}]{shimwell2017}
{Shimwell}, T.~W., {R{\"o}ttgering}, H.~J.~A., {Best}, P.~N., {et~al.} 2017,
  \aap, 598, A104

\bibitem[{{Shulevski} {et~al.}(2017){Shulevski}, {Morganti}, {Harwood},
  {Barthel}, {Jamrozy}, {Brienza}, {Brunetti}, {R{\"o}ttgering}, {Murgia},
  {White}, {Croston}, \& {Br{\"u}ggen}}]{shulevski2017}
{Shulevski}, A., {Morganti}, R., {Harwood}, J.~J., {et~al.} 2017, \aap, 600,
  A65

\bibitem[{{Shulevski} {et~al.}(2012){Shulevski}, {Morganti}, {Oosterloo}, \&
  {Struve}}]{shulevski2012}
{Shulevski}, A., {Morganti}, R., {Oosterloo}, T., \& {Struve}, C. 2012, \aap,
  545, A91

\bibitem[{{Sijacki} {et~al.}(2015){Sijacki}, {Vogelsberger}, {Genel},
  {Springel}, {Torrey}, {Snyder}, {Nelson}, \& {Hernquist}}]{sijacki2015}
{Sijacki}, D., {Vogelsberger}, M., {Genel}, S., {et~al.} 2015, \mnras, 452, 575

\bibitem[{{Slee}(1995)}]{slee1995}
{Slee}, O.~B. 1995, Australian Journal of Physics, 48, 143

\bibitem[{{Snellen} {et~al.}(2000){Snellen}, {Schilizzi}, {Miley}, {de Bruyn},
  {Bremer}, \& {R{\"o}ttgering}}]{snellen2000}
{Snellen}, I.~A.~G., {Schilizzi}, R.~T., {Miley}, G.~K., {et~al.} 2000, \mnras,
  319, 445

\bibitem[{{Soker} {et~al.}(2009){Soker}, {Sternberg}, \&
  {Pizzolato}}]{soker2009}
{Soker}, N., {Sternberg}, A., \& {Pizzolato}, F. 2009, in American Institute of
  Physics Conference Series, Vol. 1201, American Institute of Physics
  Conference Series, ed. S.~{Heinz} \& E.~{Wilcots}, 321--325

\bibitem[{{Struve} {et~al.}(2010){Struve}, {Oosterloo}, {Morganti}, \&
  {Saripalli}}]{struve2010}
{Struve}, C., {Oosterloo}, T.~A., {Morganti}, R., \& {Saripalli}, L. 2010,
  \aap, 515, A67

\bibitem[{{Subrahmanyan} {et~al.}(2006){Subrahmanyan}, {Hunstead}, {Cox}, \&
  {McIntyre}}]{subrahmanyan2006}
{Subrahmanyan}, R., {Hunstead}, R.~W., {Cox}, N.~L.~J., \& {McIntyre}, V. 2006,
  \apj, 636, 172

\bibitem[{{Swarup}(1991)}]{swarup1991}
{Swarup}, G. 1991, in Astronomical Society of the Pacific Conference Series,
  Vol.~19, IAU Colloq. 131: Radio Interferometry. Theory, Techniques, and
  Applications, ed. T.~J. {Cornwell} \& R.~A. {Perley}, 376--380

\bibitem[{{Tadhunter} {et~al.}(2011){Tadhunter}, {Holt}, {Gonz{\'a}lez
  Delgado}, {Rodr{\'{\i}}guez Zaur{\'{\i}}n}, {Villar-Mart{\'{\i}}n},
  {Morganti}, {Emonts}, {Ramos Almeida}, \& {Inskip}}]{tadhunter2011}
{Tadhunter}, C., {Holt}, J., {Gonz{\'a}lez Delgado}, R., {et~al.} 2011, \mnras,
  412, 960

\bibitem[{{Tamhane} {et~al.}(2015){Tamhane}, {Wadadekar}, {Basu}, {Singh},
  {Ishwara-Chandra}, {Beelen}, \& {Sirothia}}]{tamhane2015}
{Tamhane}, P., {Wadadekar}, Y., {Basu}, A., {et~al.} 2015, \mnras, 453, 2438

\bibitem[{{Tasse} {et~al.}(2013){Tasse}, {van der Tol}, {van Zwieten}, {van
  Diepen}, \& {Bhatnagar}}]{tasse2013}
{Tasse}, C., {van der Tol}, S., {van Zwieten}, J., {van Diepen}, G., \&
  {Bhatnagar}, S. 2013, \aap, 553, A105

\bibitem[{{Tingay} {et~al.}(2015){Tingay}, {Macquart}, {Collier}, {Rees},
  {Callingham}, {Stevens}, {Carretti}, {Wayth}, {Wong}, {Trott}, {McKinley},
  {Bernardi}, {Bowman}, {Briggs}, {Cappallo}, {Corey}, {Deshpande}, {Emrich},
  {Gaensler}, {Goeke}, {Greenhill}, {Hazelton}, {Johnston-Hollitt}, {Kaplan},
  {Kasper}, {Kratzenberg}, {Lonsdale}, {Lynch}, {McWhirter}, {Mitchell},
  {Morales}, {Morgan}, {Oberoi}, {Ord}, {Prabu}, {Rogers}, {Roshi}, {Udaya
  Shankar}, {Srivani}, {Subrahmanyan}, {Waterson}, {Webster}, {Whitney},
  {Williams}, \& {Williams}}]{tingay2015}
{Tingay}, S.~J., {Macquart}, J.-P., {Collier}, J.~D., {et~al.} 2015, \aj, 149,
  74

\bibitem[{{Turner} {et~al.}(2018){Turner}, {Rogers}, {Shabala}, \&
  {Krause}}]{turner2017}
{Turner}, R.~J., {Rogers}, J.~G., {Shabala}, S.~S., \& {Krause}, M.~G.~H. 2018,
  \mnras, 473, 4179

\bibitem[{{Turner} \& {Shabala}(2015)}]{turner2015}
{Turner}, R.~J. \& {Shabala}, S.~S. 2015, \apj, 806, 59

\bibitem[{{Vacca} {et~al.}(2018){Vacca}, {Murgia}, {Loi}, {Vazza},
  {Finoguenov}, {Carretti}, {Feretti}, {Giovannini}, {Concu}, {Melis},
  {Gheller}, {Paladino}, {Poppi}, {Valente}, {Bernardi}, {Boschin}, {Brienza},
  {Clarke}, {Colafrancesco}, {En{\ss}lin}, {Ferrari}, {de Gasperin},
  {Gastaldello}, {Girardi}, {Gregorini}, {Johnston-Hollitt}, {Junklewitz},
  {Orr{\`u}}, {Parma}, {Perley}, \& {Taylor}}]{vacca2018}
{Vacca}, V., {Murgia}, M., {Loi}, F.~G.~F., {et~al.} 2018, \mnras

\bibitem[{{van Haarlem} {et~al.}(2013){van Haarlem}, {Wise}, {Gunst}, {Heald},
  {McKean}, {Hessels}, {de Bruyn}, {Nijboer}, {Swinbank}, {Fallows},
  {Brentjens}, {Nelles}, {Beck}, {Falcke}, {Fender}, {H{\"o}randel},
  {Koopmans}, {Mann}, {Miley}, {R{\"o}ttgering}, {Stappers}, {Wijers},
  {Zaroubi}, {van den Akker}, {Alexov}, {Anderson}, {Anderson}, {van Ardenne},
  {Arts}, {Asgekar}, {Avruch}, {Batejat}, {B{\"a}hren}, {Bell}, {Bell}, {van
  Bemmel}, {Bennema}, {Bentum}, {Bernardi}, {Best}, {B{\^i}rzan}, {Bonafede},
  {Boonstra}, {Braun}, {Bregman}, {Breitling}, {van de Brink}, {Broderick},
  {Broekema}, {Brouw}, {Br{\"u}ggen}, {Butcher}, {van Cappellen}, {Ciardi},
  {Coenen}, {Conway}, {Coolen}, {Corstanje}, {Damstra}, {Davies}, {Deller},
  {Dettmar}, {van Diepen}, {Dijkstra}, {Donker}, {Doorduin}, {Dromer}, {Drost},
  {van Duin}, {Eisl{\"o}ffel}, {van Enst}, {Ferrari}, {Frieswijk}, {Gankema},
  {Garrett}, {de Gasperin}, {Gerbers}, {de Geus}, {Grie{\ss}meier}, {Grit},
  {Gruppen}, {Hamaker}, {Hassall}, {Hoeft}, {Holties}, {Horneffer}, {van der
  Horst}, {van Houwelingen}, {Huijgen}, {Iacobelli}, {Intema}, {Jackson},
  {Jelic}, {de Jong}, {Juette}, {Kant}, {Karastergiou}, {Koers}, {Kollen},
  {Kondratiev}, {Kooistra}, {Koopman}, {Koster}, {Kuniyoshi}, {Kramer},
  {Kuper}, {Lambropoulos}, {Law}, {van Leeuwen}, {Lemaitre}, {Loose}, {Maat},
  {Macario}, {Markoff}, {Masters}, {McFadden}, {McKay-Bukowski}, {Meijering},
  {Meulman}, {Mevius}, {Middelberg}, {Millenaar}, {Miller-Jones}, {Mohan},
  {Mol}, {Morawietz}, {Morganti}, {Mulcahy}, {Mulder}, {Munk}, {Nieuwenhuis},
  {van Nieuwpoort}, {Noordam}, {Norden}, {Noutsos}, {Offringa}, {Olofsson},
  {Omar}, {Orr{\'u}}, {Overeem}, {Paas}, {Pandey-Pommier}, {Pandey}, {Pizzo},
  {Polatidis}, {Rafferty}, {Rawlings}, {Reich}, {de Reijer}, {Reitsma},
  {Renting}, {Riemers}, {Rol}, {Romein}, {Roosjen}, {Ruiter}, {Scaife}, {van
  der Schaaf}, {Scheers}, {Schellart}, {Schoenmakers}, {Schoonderbeek},
  {Serylak}, {Shulevski}, {Sluman}, {Smirnov}, {Sobey}, {Spreeuw}, {Steinmetz},
  {Sterks}, {Stiepel}, {Stuurwold}, {Tagger}, {Tang}, {Tasse}, {Thomas},
  {Thoudam}, {Toribio}, {van der Tol}, {Usov}, {van Veelen}, {van der Veen},
  {ter Veen}, {Verbiest}, {Vermeulen}, {Vermaas}, {Vocks}, {Vogt}, {de Vos},
  {van der Wal}, {van Weeren}, {Weggemans}, {Weltevrede}, {White}, {Wijnholds},
  {Wilhelmsson}, {Wucknitz}, {Yatawatta}, {Zarka}, {Zensus}, \& {van
  Zwieten}}]{vanhaarlem2013}
{van Haarlem}, M.~P., {Wise}, M.~W., {Gunst}, A.~W., {et~al.} 2013, \aap, 556,
  A2

\bibitem[{{van Weeren} {et~al.}(2016){van Weeren}, {Williams}, {Hardcastle},
  {Shimwell}, {Rafferty}, {Sabater}, {Heald}, {Sridhar}, {Dijkema}, {Brunetti},
  {Br{\"u}ggen}, {Andrade-Santos}, {Ogrean}, {R{\"o}ttgering}, {Dawson},
  {Forman}, {de Gasperin}, {Jones}, {Miley}, {Rudnick}, {Sarazin}, {Bonafede},
  {Best}, {B{\^i}rzan}, {Cassano}, {Chy{\.z}y}, {Croston}, {Ensslin},
  {Ferrari}, {Hoeft}, {Horellou}, {Jarvis}, {Kraft}, {Mevius}, {Intema},
  {Murray}, {Orr{\'u}}, {Pizzo}, {Simionescu}, {Stroe}, {van der Tol}, \&
  {White}}]{vanweeren2016}
{van Weeren}, R.~J., {Williams}, W.~L., {Hardcastle}, M.~J., {et~al.} 2016,
  \apjs, 223, 2

\bibitem[{{Venturi} {et~al.}(2004){Venturi}, {Dallacasa}, \&
  {Stefanachi}}]{venturi2004}
{Venturi}, T., {Dallacasa}, D., \& {Stefanachi}, F. 2004, \aap, 422, 515

\bibitem[{{Wagner} \& {Bicknell}(2011)}]{wagner2011}
{Wagner}, A.~Y. \& {Bicknell}, G.~V. 2011, \apj, 728, 29

\bibitem[{{Wagner} {et~al.}(2012){Wagner}, {Bicknell}, \&
  {Umemura}}]{wagner2012}
{Wagner}, A.~Y., {Bicknell}, G.~V., \& {Umemura}, M. 2012, \apj, 757, 136

\bibitem[{{Wegner} {et~al.}(1993){Wegner}, {Haynes}, \&
  {Giovanelli}}]{wegner1993}
{Wegner}, G., {Haynes}, M.~P., \& {Giovanelli}, R. 1993, \aj, 105, 1251

\bibitem[{{White} \& {Becker}(1992)}]{white1992}
{White}, R.~L. \& {Becker}, R.~H. 1992, \apjs, 79, 331

\bibitem[{{Williams} {et~al.}(2016){Williams}, {van Weeren}, {R{\"o}ttgering},
  {Best}, {Dijkema}, {de Gasperin}, {Hardcastle}, {Heald}, {Prandoni},
  {Sabater}, {Shimwell}, {Tasse}, {van Bemmel}, {Br{\"u}ggen}, {Brunetti},
  {Conway}, {En{\ss}lin}, {Engels}, {Falcke}, {Ferrari}, {Haverkorn},
  {Jackson}, {Jarvis}, {Kapi{\'n}ska}, {Mahony}, {Miley}, {Morabito},
  {Morganti}, {Orr{\'u}}, {Retana-Montenegro}, {Sridhar}, {Toribio}, {White},
  {Wise}, \& {Zwart}}]{williams2016}
{Williams}, W.~L., {van Weeren}, R.~J., {R{\"o}ttgering}, H.~J.~A., {et~al.}
  2016, \mnras, 460, 2385

\bibitem[{{Worrall} \& {Birkinshaw}(2006)}]{worrall2006}
{Worrall}, D.~M. \& {Birkinshaw}, M. 2006, in Lecture Notes in Physics, Berlin
  Springer Verlag, Vol. 693, Physics of Active Galactic Nuclei at all Scales,
  ed. D.~{Alloin}, 39

\bibitem[{{Wu}(2009)}]{wu2009}
{Wu}, Q. 2009, \apjl, 701, L95

\bibitem[{{Zhao} {et~al.}(1993){Zhao}, {Sumi}, {Burns}, \& {Duric}}]{zhao1993}
{Zhao}, J.-H., {Sumi}, D.~M., {Burns}, J.~O., \& {Duric}, N. 1993, \apj, 416,
  51

\end{thebibliography}

\end{document}